\documentclass[a4paper,11pt]{article}
\pdfoutput=1
\usepackage{jcappub}
\usepackage[T1]{fontenc}
\usepackage{float}
\usepackage{array}
\usepackage{graphicx}
\usepackage{xcolor}  
\usepackage{hyperref}
\usepackage{siunitx}
\usepackage{aas_macros}
\usepackage{booktabs}

\newcolumntype{M}[1]{>{\centering\arraybackslash$}p{#1}<{$}}

\title{The Role of Big Bang Nucleosynthesis in Joint Cosmological Analyses}

\author[a,b,1]{Mehmet Akharman,\note{Corresponding author.}}
\author[a]{Nicholas DePorzio,}
\author[c,d]{Cara Giovanetti,}
\author[a]{and Hongwan Liu}

\affiliation[a]{Physics Department, Boston University,\\ Boston, MA 02215, USA}
\affiliation[b]{Physics Department, University of California, Santa Barbara,\\ Santa Barbara, CA 93106, USA}
\affiliation[c]{Theory Group, Lawrence Berkeley National Laboratory,\\ Berkeley, CA 94720, USA}
\affiliation[d]{Leinweber Institute for Theoretical Physics, University of California,\\ Berkeley, CA 94720, USA}

\emailAdd{akharman@ucsb.edu}
\emailAdd{deporzio@bu.edu}
\emailAdd{cgiovanetti@lbl.gov}
\emailAdd{hongwan@bu.edu}

\abstract{We perform joint Baryon Acoustic Oscillation (BAO) + Big Bang Nucleosynthesis (BBN) and BAO+BBN+Cosmic Microwave Background (CMB) analyses with Planck CMB and DESI DR2 BAO data, explicitly marginalizing over BBN nuisance parameters for the first time with these data combinations and paying particular attention to the impact of BBN on these results.  We find in our fiducial analyses $h=0.6833^{+0.0048}_{-0.0053}$ ($\Lambda$CDM, BAO+BBN), $h=0.6823^{+0.0027}_{-0.0026}$ ($\Lambda$CDM, BAO+BBN+CMB), as well as $h=0.6791^{+0.0069}_{-0.0075},N_{\rm{eff}}=2.974^{+0.098}_{-0.099}$ ($\Lambda$CDM+$N_{\rm{eff}}$, BAO+BBN), and $h=0.6838^{+0.0052}_{-0.0054},N_{\rm{eff}}=3.065^{+0.078}_{-0.076}$ ($\Lambda$CDM+$N_{\rm{eff}}$, BAO+\\BBN+CMB). We demonstrate how investigator choices can impact results of joint analyses involving BBN.  We provide recommendations for the treatment of BBN in light of recent work, and demonstrate a pipeline that accurately accounts for prediction uncertainties in BBN.}

\begin{document}
\maketitle
\flushbottom

\section{Introduction}\label{sec:intro}
The era of precision cosmology has brought high-quality data across a range of epochs, observables, and scales. Cosmological models must demonstrate consistency with not just the Cosmic Microwave Background (CMB) anisotropy power spectrum, a longstanding benchmark, but also with data from Baryon Acoustic Oscillations (BAO) and Big Bang Nucleosynthesis (BBN). 
These latter measurements are able to constrain cosmological parameters at comparable precision to the CMB~\cite{Cooke_2018,aver_2026,Adame_2025}.

The Lambda Cold Dark Matter ($\Lambda$CDM) paradigm requires consistency between each of these epochs.  Of the six parameters that define the $\Lambda$CDM fit to the CMB data, BBN is directly and independently sensitive to one: the baryon abundance, $\Omega_bh^2$ (hereinafter $\omega_b$). With prior information on the CMB temperature $T_\gamma$ and from BBN on $\omega_b$, BAO data can also be used to infer the late-time matter density $\Omega_m$, and the Hubble constant $H_0$. 
All three observables are also sensitive to the effective number of neutrino species $N_{\rm eff}$, if it is treated as a free parameter.  
These three probes and their various combinations should provide a consistent determination of the parameters for which their sensitivities overlap.

Many previous analyses have evaluated these datasets for $\Lambda$CDM consistency.  By and large, $\Lambda$CDM seems to provide a somewhat consistent explanation for the parameter values recovered by analyses of the CMB, BAO, and BBN, though discrepancies as large as $2-3\sigma$ routinely appear (for example, the ACT preference for a low $N_{\rm{eff}}$~\cite{calabrese2025atacamacosmologytelescopedr6}, the $hr_d$~\cite{DESI:2025zgx} and $n_s$~\cite{ferreira2025baocmbtensionimplicationsinflation} tensions between BAO and the CMB, or the baryon density discrepancy between the CMB and BBN~\cite{Pitrou_2021,launders2026,Poulin:2026ltf}). These small discrepancies may be some combination of expected statistical fluctuations, hints for physics beyond $\Lambda$CDM, or artifacts of untracked experimental and analysis systematics.  Careful control over analysis methods and choices can help to eliminate this last category as an explanation for these unexpected results.

BBN has matured as a precision science relatively recently, with results of the primordial deuterium abundance from 2018 reaching percent-level precision~\cite{Cooke_2018} and measurements of the primordial helium-4 abundance reaching sub-percent precision in 2026~\cite{aver_2026}.  These measurements provide a correspondingly tight determination of the baryon density and $N_{\rm{eff}}$, given the high sensitivity of these abundances to these parameters.  Indeed, BBN is currently the most sensitive independent probe of $N_{\rm{eff}}$~\cite{yeh2026}.  However, taking advantage of this information  requires some attention to detail, and differing analysis choices or treatments of uncertainties can significantly impact results. For example, different choices of the astrophysical measurements of primordial abundances used in constructing a BBN likelihood can change the inference of the baryon density (see e.g.\ ref.~\cite{schoneberg_2024}).  

There are uncertainties related to the prediction of the abundances given a model, dominated primarily by uncertainties in the nuclear reaction rates (especially $d(d,p)t$, $d(d,n)^3$He, and $d(p,\gamma)^3$He~\cite{Pitrou_2021,Pisanti_2021,Yeh_2021}), and varying treatments of these uncertainties by different teams of investigators can impact BBN predictions.  
Two important choices in particular can lead to significantly different outcomes in parameter inference.
First, the central value and uncertainty of these rates are estimated by performing a fit to laboratory measurements of nuclear reaction cross sections (or, equivalently, astrophysical $S$-factors; see discussion in e.g.\ ref.~\cite{Pitrou_2018}); depending on the details of the fit procedure, different groups obtain reaction rates that predict significantly discrepant deuterium abundances,\footnote{Note that helium-4 abundance predictions are usually in agreement between groups, within experimental uncertainties.  The helium-4 abundance is primarily determined by electroweak physics at the nucleon level, which is under much better theoretical control than nuclear reaction rates.} as will be discussed in detail in section~\ref{sec:bbn_like}.  The effect of this systematic is appreciated in the BBN literature (see e.g.\ discussion in refs.~\cite{Pisanti_2021,Pitrou_2021,launders2026}), but is not always explored in joint analyses.  For example, the DESI DR2 analysis uses a prior on the baryon density from just one reaction network~\cite{Adame_2025}. SPT discusses only the effects of reaction network systematics on helium-4 predictions, reporting accurately that systematic effects are negligible there, and eschews the network-dependent BBN constraint on $\omega_b$ from deuterium entirely~\cite{Camphuis_2026}.  We will demonstrate below that using other reaction networks---including the fiducial network that we recommend---can change the inferred cosmological parameters.

Second, once a reaction network has been chosen, the rate uncertainties must still be incorporated into a BBN likelihood. 
These uncertainties are often captured by a single nuisance parameter which rescales the rate in a temperature-dependent manner, with a prior Gaussian distribution~\cite{Pitrou_2018,Burns_2023}; more recent work models these rates as Gaussian processes that capture these uncertainties in a more flexible manner~\cite{launders2026}.
In a careful treatment of these uncertainties, these nuisance parameters should be included in the likelihood, and marginalized over to properly infer cosmological parameters (see e.g.\ refs.~\cite{Pitrou_2018,Burns_2023,LINX_short}). 
Many analyses, however, do not follow this procedure, and instead rely on compressed likelihoods for $\Lambda$CDM and $\Lambda$CDM+$N_{\rm eff}$. 
Results from the Planck mission~\cite{Planck2018}, for example, rely on an estimate of uncertainties on the deuterium prediction by varying all reaction rates simultaneously one standard deviation high and one standard deviation low, which overestimates BBN uncertainties~\cite{LINX_short}.  Then, that estimated uncertainty is added to a joint likelihood as a constant, despite the true prediction uncertainty varying as a function of cosmological model parameters $\omega_b$ and $N_{\rm{eff}}$.  ACT performs a similar procedure, using a constant estimate of the combined theory and observational uncertainty over their whole parameter range~\cite{calabrese2025atacamacosmologytelescopedr6}.  As explored in ref.~\cite{LINX_short}, these choices are less impactful in combination with the CMB than in BBN alone in $\Lambda$CDM.  
However, it is unclear if compressed likelihoods are still suitable as BBN precision continues to increase. 
They are also currently unavailable for any other new physics model beyond $N_{\rm eff}$, require high-dimensional parameter scans to assemble, and have to be updated whenever a new reaction rate is published (e.g.\  those computed using effective field theories in refs.~\cite{Tait:2026guk,Tait:2026vrd}).

In this work, we use a joint BAO+BBN likelihood to present new results on a CMB-independent determination of $h$ (following on the work of e.g.~refs.~\cite{Addison_2018,Cuceu_2019,Schoneberg_2019}) using DESI DR2 data~\cite{DESI:2025zgx} in $\Lambda$CDM and $\Lambda$CDM+$N_{\rm eff}$, the latter of which has not been done with DESI data, as far as the authors are aware.  We also present a BAO+BBN+CMB analysis by including Planck data of both $\Lambda$CDM and $\Lambda$CDM+$N_{\rm eff}$. 
Combined analyses of this type have been performed in the literature; most recently, ref.~\cite{goldstein20262determinationnrmeff} reevaluated bounds on $N_{\rm eff}$ given the recent results in ref.~\cite{aver_2026}. 
Here, we show for the first time results of a BAO+BBN+CMB analysis using a full BBN likelihood, marginalizing over all BBN nuclear rate uncertainties with no reliance on existing BBN posteriors in the literature to use as a prior in our joint analyses. 
We also carefully study systematic uncertainties associated with different choices of nuclear reaction networks.

The BAO+BBN+CMB data combination was not carried out in full in DESI DR2, apart from a BAO+BBN analysis with a prior on $\theta_*$ (the angular size of the sound horizon at the epoch of last scattering) to summarize limited CMB information~\cite{DESI:2025zgx}.  Since BAO is sensitive to the combination $H_0r_d$, where $r_d$ is the comoving sound horizon at the end of the drag epoch, another dataset must be included to constrain $\omega_b$ and break this degeneracy to attain sensitivity to $H_0$.  From the perspective of a BAO analysis, this additional information may come from a joint analysis with the CMB, where the CMB determination of $\omega_b$ is the most precise available, or it may come from BBN, so that the joint analysis is independent of the CMB power spectrum measurements.  With this framing, including BBN information in a BAO+CMB analysis may seem redundant.  However, this neglects the facts that \textit{1)} BBN does provide significant information on $\omega_b$, and so a full combination of data should be used; and \textit{2)} BBN provides strong constraints on $N_{\rm{eff}}$, \textit{correlated with its $\omega_b$ constraint}; a joint analysis of this kind would take advantage of CMB precision for $\omega_b$ and BBN precision for $N_{\rm{eff}}$, improving the precision with which $N_{\rm eff}$ can currently be inferred with BAO data, e.g.\ in the CMB+DESI data analysis in ref.~\cite{Elbers:2025vlz}. 

Along the way we highlight the impacts of our BBN analysis choices and provide recommendations for researchers performing joint BAO, BBN, and CMB analyses.  We illustrate the effects of these choices in both $\Lambda$CDM and $\Lambda$CDM+$N_{\rm{eff}}$ cosmologies.   
We compare our full joint results across reaction networks and across strategies for including constraints from BBN via summaries like priors.  We provide critical validation of literature results that use summaries rather than the full likelihood.  However, we caution that not all summary procedures are equivalent, and results claiming high precision may still require the full likelihood treatment; we suggest analyses that require the precision from marginalizing over CMB nuisance parameters should also marginalize over BBN nuisance parameters, rather than relying on pre-marginalized likelihoods.  We also emphasize that \textbf{the determination of $N_{\rm{eff}}$ in the joint analysis depends on the choice of reaction network}, an effect illustrated in ref.~\cite{LINX_short} but not taken into account in more recent analyses~\cite{yeh2026,goldstein20262determinationnrmeff}.

The rest of the paper is organized as follows.  In the next section, we provide a detailed overview of the methodology used to incorporate each observable component into our analyses.  We also provide a more detailed discussion of BBN systematic uncertainties and our recommendations for their treatment.  Then, we present results from different combinations of these three datasets, beginning with BAO+BBN in section~\ref{sec:BAO_BBN_no_Neff} and moving to BAO+BBN+CMB in section~\ref{sec:all3}.  In the latter section, we also discuss important parameter relationships, effects on BBN nuisance parameters, and consequences of varying treatments of BBN in these analyses.  We summarize our recommendations for the treatment of BBN in joint analyses and discuss the implications of our results in section~\ref{sec:discussion}.

\section{Likelihoods and methods}

All analyses in this work use some combination of BAO, BBN, and CMB data.  The parameters, theory predictions, datasets, and likelihoods used in each are described in the subsections below. 

For each analysis in $\Lambda$CDM and $\Lambda$CDM+$N_{\rm{eff}}$, we use nested sampling~\cite{Skilling_2004,Skilling_2006} to estimate the model and nuisance parameters preferred by each combination of probes.  We use \texttt{dynesty}~\cite{Speagle_2020,speagle_2023} with a static sampler~\cite{Feroz_2009,Neal_2003,Handley_2015a,Handley_2015b} and 500 live points, terminating sampling once we achieve an evidence tolerance of $\Delta \ln Z = 0.5$ for evidence $Z$.

The total log-likelihood is the sum of individual log-likelihood terms, maximally
\begin{equation}
\label{eq:lltot}
    \ln\mathcal{L}_{\rm tot} = \ln\mathcal{L}_{\rm BBN} + \ln\mathcal{L}_{\rm BAO} + \ln\mathcal{L}_{\rm CMB}
\end{equation}
with e.g.\ $\ln\mathcal{L}_{\rm CMB}$ excluded in BAO+BBN analyses.  We define each of these terms in the subsections that follow.

We use wide, uniform priors for our model parameters, with ranges that mirror choices commonly made in the literature. For $\omega_b$ and $N_\text{eff}$, we set the ranges to those used in ref.~\cite{LINXlong} (in $\Lambda$CDM, $N_{\rm{eff}}$ is fixed to 3.044 as per refs.~\cite{Akita:2020szl,Froustey:2020mcq,Bennett:2020zkv}). Our priors for $h$ and $\Omega_m$ follow those in Table 2 of ref.~\cite{Adame_2025}, and those for $A_s$, $n_s$, and $\tau_{\rm{reio}}$ follow ref.~\cite{Planck_2018_likelihoods}. The prior ranges are displayed in Table~\ref{tab:priors}. 

\begin{table}[t]
\centering
\begin{tabular}{lc}
\toprule
Parameter & Prior \\
\midrule
$100\omega_b$ & $\mathcal{U}[1.9,\;2.5]$ \\
$\Omega_m$ & $\mathcal{U}[0.01,\;0.99]$ \\
$h$ & $\mathcal{U}[0.2,\;1.0]$ \\
$N_{\mathrm{eff}}$ & $\mathcal{U}[1.15,\;4.96]$ \\
$10^{9}A_s$ & $\mathcal{U}[1,\;3]$ \\
$n_s$ & $\mathcal{U}[0.85,\;1.0]$ \\
$100\tau_{\mathrm{reio}}$ & $\mathcal{U}[2.5,\;9]$ \\
\bottomrule
\end{tabular}
\caption{Uniform priors adopted for the sampled model parameters.  The $N_{\rm{eff}}$ prior is only applicable in $\Lambda$CDM+$N_{\rm{eff}}$ analyses.  It is formally a prior on the LINX parameter \texttt{Delta\_Neff\_init} (see ref.~\cite{LINXlong}) in the range [-7,7], though we translate that (bijective) range to the equivalent range in $N_{\rm{eff}}$ for ease of interpretation.}
\label{tab:priors}
\end{table}

\subsection{BBN likelihood and systematics}\label{sec:bbn_like}

As discussed in section~\ref{sec:intro}, different groups have advocated the use of different reaction networks, which are derived from fits to experimentally-measured cross sections ($S$-factors) for each nuclear reaction rate.  Many choices of reaction network exist in the literature; among the most common are that used by the public BBN code PRIMAT~\cite{Pitrou_2018}, that used in the code PArthENoPE~\cite{Consiglio_2018,Gariazzo:2021iiu,Pisanti_2021}, and the PRyMordial~\cite{burns2023prymordialminutesstandardmodel} reaction network similar to that used in ref.~\cite{Yeh_2021} using data from the NACREII database~\cite{Xu_2013}, which we denote ``PRyM/YOF.''  

There are two axes of disagreement among these different reaction networks: dataset selection, and fitting or interpolation procedure.  There are many measurements of the cross sections for the nuclear reaction rates contributing to the primordial deuterium prediction; each group makes its own determination of which datasets to include and exclude.  The PRIMAT reaction network is the most selective in its dataset curation, while the PArthENoPE and PRyM/YOF networks include roughly the same number of datasets (though they do not overlap entirely in the datasets they include). 

In this paper, we reemphasize some of the recommendations made in ref.~\cite{launders2026}, which also justify the choice of our fiducial network here. 
For a given reaction, we do not recommend the use of datasets that fall far outside the reaction's energy range relevant for BBN (0.015-0.6 \si{MeV} for the rates most important for determining the primordial deuterium abundance~\cite{launders2026}).  Including such a dataset risks introducing systematics that can bias the fit to data at BBN-relevant energies.  This criterion excludes, for example, the dataset in ref.~\cite{Schulte1972}, whose lowest energy datum lies at \SI{1.96}{MeV}; this dataset is included in the PArthENoPE and PRyM/YOF networks, but excluded in the PRIMAT network.\footnote{The reasoning for excluding this rate in ref.~\cite{Gomez_Inesta_2017} and therefore the PRIMAT network is a reluctance to allow \textit{extrapolation of the theory prediction} of ref.~\cite{Arai2011} to high energies to bias the overall fit.  Our concerns are related more closely to the potential introduction of inconsistencies or systematics at high energy, similar to concerns noted in ref.~\cite{Coc_2015} for the data from ref.~\cite{Tumino2014}.}  In addition, the dataset in ref.~\cite{Tumino2014} uses a measurement procedure that is not directly sensitive to the overall normalization of the measured cross section, and instead the overall normalization is obtained using information from other datasets, many of which are also used on their own in these reaction networks.  As noted in ref.~\cite{Pitrou_2021}, this dataset is therefore not independent of the others and its errors and covariance must be weighted carefully if it is to be used in determining a rate in a BBN reaction network, though there is not enough information provided in ref.~\cite{Tumino2014} to extract these covariances.  We therefore do not recommend the use of this dataset either; it is used in the PArthENoPE and PRyM/YOF networks, and excluded in the PRIMAT network.

Nuclear reaction $S$-factors must be thermally averaged to obtain their rates as a function of temperature; this is frequently done by fitting a functional form to the data. This fitting procedure is the other primary root of disagreement between reaction networks.  The PRIMAT reaction network scales the theory predictions of ref.~\cite{Arai2011} to fit the data~\cite{Coc_2015,Gomez_Inesta_2017,Moscoso_2021}; the shape of the fitted function is relatively inflexible.  The PArthENoPE reaction network uses low-degree polynomials to fit the data~\cite{Pisanti_2021}.  The PRyM/YOF network uses potential models to incorporate theory information in a manner more flexible than the \textit{ab initio} theory curve used in the PRIMAT network for the relevant deuterium fusion rates from the NACREII database, along with the polynomial fit for the rate $d(p,\gamma)^3$He from ref.~\cite{Mossa_2020}.\footnote{The PRyM/YOF network was originally compiled by the authors of the PRyMordial BBN code~\cite{burns2023prymordialminutesstandardmodel}, though the LINX version drops two $^7$Be processes and adds $t(p,\gamma)^4$He (see ref.~\cite{LINXlong}).  The original PRyMordial version of this network uses the same rates as Yeh, Olive, and Fields in ref.~\cite{Yeh_2021} (hence our use of the acronym ``YOF''), apart from $d(p,\gamma)^3$He.  The PRyMordial code instead uses the published LUNA fit to $d(p,\gamma)^3$He data~\cite{Mossa_2020} while the authors of ref.~\cite{Yeh_2021} perform their own polynomial fit to these data.  The authors of ref.~\cite{Yeh_2021} and colleagues have also updated their network to include their own polynomial fits for $d(d,n)^3$He and $d(d,p)^3$H in ref.~\cite{Yeh_2022}, a change that has not been propagated to the LINX ``key\_YOF'' network (the reaction network used here), or, to our knowledge, the PRyMordial ``key\_nacreii\_rates'' (the equivalent network in PRyMordial from which the LINX network is derived).}  

Ref.~\cite{launders2026} recently demonstrated that the polynomial fitting procedure used by the PArthENoPE network produces a bias in the inferred value for primordial deuterium.  This reference also used a validated Gaussian process regression procedure for interpolation and recovered a primordial deuterium abundance with a central value and uncertainty both in agreement with the prediction from PRIMAT.  In light of these results, we use the PRIMAT network for our fiducial analyses.  We include results using other networks in this work to better contextualize existing literature analyses, to illustrate the impact of changing relative error bars between BBN and the CMB, and to illustrate the effects of introducing or alleviating tension between BBN and the CMB.

The choice of reaction network determines the predicted primordial element abundances that appear in the BBN likelihood.  All BBN analyses in this work use LINX~\cite{LINXlong} for theory predictions, with the ``key'' versions of the aforementioned reaction networks provided in LINX.  These truncated reaction networks, containing only 12 nuclear reactions, are sufficient for accurate predictions of both the primordial deuterium abundance relative to hydrogen, D/H, and the proxy for the primordial helium abundance\footnote{While this is conventionally referred to as the ``helium-4 mass fraction'' and denoted $\textrm{Y}_{\textrm{P}}$ in the BBN literature, we instead reserve that notation for the true helium-4 mass fraction used in CMB analyses, discussed below.} $\textrm{Y}_{\textrm{P}}^{\rm{BBN}}=4n_{\rm{He}}/n_{\rm{B}}$~\cite{smith_1993,Nollett_2000,Serpico_2004,coc_2010,Cyburt_2016,Fields_2020,LINXlong}, where $n_{\rm{He}}$ and $n_{\rm{B}}$ are the helium and baryon number densities, respectively.  This introduces 13 nuisance parameters into our BBN analyses: one for each of the amplitudes of the 12 nuclear reaction rates, and one sampling the uncertainty in the neutron lifetime.  These nuisance parameters and their sampling are discussed in more detail in ref.~\cite{LINXlong}; we use the same $\mathcal{N}(0,1)$ prior for the rate nuisance parameters and sample from a Gaussian prior $\tau_n = 879.4^{+0.6}_{-0.6}~\SI{}{\second}$ for the neutron lifetime~\cite{Czarnecki:2018okw}.

The only relevant model parameter for BBN analyses in $\Lambda$CDM is the baryon density $\omega_b$.  When $N_{\rm{eff}}$ is allowed to vary in $\Lambda$CDM+$N_{\rm{eff}}$, it also factors into the BBN likelihood. Given input model and nuisance parameters, and a choice of reaction network, LINX outputs D/H and $\textrm{Y}_{\textrm{P}}^{\rm{BBN}}$. 

These outputs are compared against measurement in the likelihood.  We use 
\begin{alignat*}{2}
    \textrm{D/H}^{\rm{obs}} &= 2.527\times10^{-5}\hspace{1cm} 
    \sigma_{\textrm{D/H}^{\rm{obs}}} &&= 0.030\times10^{-5}\\
    \textrm{Y}_{\textrm{P}}^{\rm{BBN,obs}} &=0.2458  \hspace{2cm}
    \sigma_{\textrm{Y}_{\textrm{P}}^{\rm{BBN,obs}}} &&=0.0013 \,,
\end{alignat*}
for the observed primordial abundances, where the deuterium abundance is taken from ref.~\cite{Cooke_2018} and the helium-4 abundance is taken from ref.~\cite{aver_2026}.  We select ref.~\cite{Cooke_2018} for the deuterium abundance as it results from a reanalysis of several high-quality astrophysical systems in one self-consistent analysis, which is a more principled treatment than a simple average over many results that may not be consistent in their analysis choices.  We use the helium-4 abundance from ref.~\cite{aver_2026}, a significantly improved measurement compared to previous work.  The unprecedented precision of this result is achieved in part by eliminating a systematic uncertainty incurred by regressing helium abundance measurements inferred from spectra at finite metallicity to zero metallicity.  This analysis instead uses a weighted average of the lowest-metallicity astrophysical systems observed in this campaign to estimate the primordial helium-4 abundance.  This procedure appropriately weights the lowest-metallicity systems, rather than allowing high-metallicity systems to unduly influence the inference for the primordial helium-4 abundance as in the case where linear regression is used.

The overall BBN likelihood is finally given by
\begin{equation}
\label{eq:llBBN}
\ln\mathcal{L}_{\rm{BBN}} = -\frac{1}{2}\left[\left(\frac{\textrm{Y}_{\textrm{P}}^{\rm{BBN}}-\textrm{Y}_{\textrm{P}}^{\rm{BBN,obs}}}{\sigma_{\textrm{Y}_{\textrm{P}}^{\rm{BBN,obs}}}}\right)^{2}+\left(\frac{\textrm{D/H}-\textrm{D/H}^{\rm{obs}}}{\sigma_{\textrm{D/H}^{\rm{obs}}}}\right)^{2}\right].
\end{equation}

\subsection{BAO likelihood}

As recombination proceeded and the early universe became neutral, the Compton drag between the photon and baryon fluids decreased dramatically, becoming negligible at the end of the drag epoch at redshift $z_d$. 
At this point, acoustic waves in the baryon fluid essentially stopped propagating, ultimately leading to an enhanced clustering of matter on comoving length scales comparable to the comoving sound horizon at $z_d$,  
\begin{equation}
r_d \equiv \int_{z_d}^\infty dz' \frac{c_s(z')}{H(z')} \,.
\label{eq:rd_bao}
\end{equation}
Here, $c_s$ is the sound speed of the photon-baryon fluid,
\begin{equation}
    c_s(z) = \frac{c}{\sqrt{3(1+R(z))}} \,,
\end{equation}
with deviations from $c_s = c/\sqrt{3}$, where $c$ is the speed of light,   determined entirely by the baryon loading factor,
\begin{equation}
    R(z) \equiv \frac{3\rho_b(z)}{4\rho_\gamma(z)} = \frac{3\omega_b}{4\omega_\gamma}\frac{1}{1+z} \,.
\end{equation}
During the drag epoch, the Hubble parameter is given by 
\begin{equation}
H^2(z) = \left(100~{\rm km\,s^{-1}\,Mpc^{-1}}\right)^2 \left\{ \omega_{bc}(1+z)^3 + \omega_\gamma\left[1 + \frac{7}{8}\left(\frac{4}{11}\right)^{4/3} N_{\rm eff}\right](1+z)^4 \right\}, 
\label{eq:H_bao}
\end{equation}
where $\omega_{bc}$ is the combined energy density of baryons and CDM. 
The Far-Infrared Absolute Spectrophotometer (FIRAS) measurement of the CMB temperature $T_\gamma$~\cite{Fixsen_2009} provides an extremely precise measurement of $\omega_\gamma \propto T_\gamma^4$, and so ultimately $r_d$ is directly dependent on only $\omega_b$ and $\omega_{bc}$, as well as $N_{\rm eff}$ in $\Lambda$CDM+$N_{\rm eff}$. 

In a flat Universe, measuring the angular scale of the BAO feature at redshift $z$ provides a measurement of the ratio of $r_d$ to the transverse comoving distance $D_M(z)$, given by
\begin{equation}
D_M(z) \equiv c \int_0^z \frac{dz'}{H(z')}.\label{eq:DM}
\end{equation}
On the other hand, determining the BAO feature along the line-of-sight in redshift gives us the ratio of $r_d$ to the Hubble distance,
\begin{equation}
D_H(z) \equiv \frac{c}{H(z)}.
\end{equation}
Another useful combined quantity reported by DESI is the volume-average of the two distances,
\begin{equation}
D_V(z) \equiv \left[ D_M^2(z)\, D_H(z)\, z \right]^{1/3}.
\end{equation}
At late times, the Hubble parameter in $\Lambda$CDM is given by
\begin{alignat}{1}
    H(z) = \left(100 h~{\rm km\,s^{-1}\,Mpc^{-1}}\right) \sqrt{ \Omega_m(1+z)^3 + 1 - \Omega_m},
\end{alignat}
with the contribution from radiation assumed to be negligible. 
From this expression, we see that $D_M/r_d$, $D_H/r_d$ and $D_V/r_d$---the quantities that are directly measured---are all only dependent on the parameter combination $hr_d$ and on $\Omega_m$. 
The uncalibrated BAO measurement---with the degeneracy between $h$ and $r_d$ left unbroken---is therefore able to determine $\Omega_m$ without any additional input from other experiments. 

If we further calibrate the BAO measurement to a value of $r_d$, e.g.\ by using a value predicted by $\Lambda$CDM, BAO can then be separately sensitive to both $h$ and $\Omega_m$. 
Given the parameter dependence of $r_d$ on $\omega_b$, $\omega_{bc}$ and $N_{\rm eff}$, the BAO measurement can be calibrated by \textit{1)} specifying the assumptions on the mass of the neutrino species, which establishes a relation between $\Omega_m$, $h$ and $\omega_{bc}$ (see ref.~\cite{DESI:2025zgx} for a thorough discussion), and \textit{2)} providing information on $\omega_b$ and $N_{\rm eff}$ from the CMB, BBN, or both.
These steps are then sufficient to predict $r_d$ given cosmological parameters $\Omega_m$, $\omega_b$, $h$, as well as $N_{\rm eff}$ where applicable, allowing for inference of their values conditioned on the chosen combination of datasets.
Most notably, the combination of BAO+BBN can provide an inference of $h$ independent of CMB anisotropies, a useful cross-check of CMB systematics for the Hubble tension (see e.g.\ refs.~\cite{CosmoVerseNetwork:2025alb, Shah:2021onj}). 

Throughout this paper, we assume one massive neutrino species with a mass of $m_\nu = \SI{0.06}{eV}$.  For a given input vector of cosmological parameters, we use CLASS\footnote{See \url{https://github.com/lesgourg/class_public}}~\cite{lesgourgues_2011a,Blas:2011rf,lesgourgues_2011c,lesgourgues_2011d} to compute $r_d$.\footnote{As per the CLASS documentation, we reduce the CLASS \texttt{N\_ur} parameter by 1.0132 from our input $N_{\rm{eff}}$ (either 3.044 in $\Lambda$CDM or a value within the prior in table~\ref{tab:priors}) to account for the contribution of the massive neutrino to the energy density in radiation at early times.  For a recent discussion of the treatment of the massive neutrino distribution that leads to this convention, see ref.~\cite{abcmb}.}  
We then integrate Eq.~\eqref{eq:DM} directly using trapezoidal integration over a finely sampled redshift grid (10,000 linearly spaced points between $z=0$ and the target redshift) to construct the predicted ratios $D_M/r_d$, $D_H/r_d$, and $D_V/r_d$. 
In computing these ratios, we track the energy density of the massive neutrino by integrating over its Fermi-Dirac distribution function.

We then compare these predictions to the DESI DR2 BAO measurements, in the form of distance ratios $D_M(z)/r_d$, $D_H(z)/r_d$, and $D_V(z)/r_d$ for each tracer (BGS, LRG, ELG, QSO, Ly$\alpha$) at the redshifts provided in Table IV of the DESI DR2 data release spanning $0.295\leq z \leq 2.330$~\cite{DESI:2025zgx}. Note that these values are fully marginalized over the BAO nuisance parameters, including Redshift Space Distortion effects ~\cite{DESI:2025qqy}, and the exact values used are those provided by the Cobaya sampler\footnote{See \url{https://github.com/CobayaSampler/bao_data/blob/master/desi_bao_dr2/desi_gaussian_bao_ALL_GCcomb_mean.txt}}~\cite{Torrado:2020dgo, 2019ascl.soft10019T} (though this analysis does not use Cobaya). 

We use these predictions and data to construct a BAO likelihood assuming a multivariate Gaussian distribution:
\begin{equation}
\chi^2 = (\mathbf{m} - \mathbf{d})^{\mathrm T} C^{-1} (\mathbf{m} - \mathbf{d}),
\end{equation}
where $\mathbf{m}$ is the vector of model predictions, $\mathbf{d}$ is the vector of DESI DR2 measurements, and $C$ is the published DESI DR2 covariance matrix\footnote{See \url{https://github.com/CobayaSampler/bao_data/blob/master/desi_bao_dr2/desi_gaussian_bao_ALL_GCcomb_cov.txt}}~\cite{DESI:2025zpo, DESI:2025zgx, DESI:2025qqy}. The log-likelihood contribution is finally
\begin{equation}
\ln \mathcal{L}_{\mathrm{BAO}} = -\frac{1}{2}\chi^2.
\end{equation}

\subsection{CMB likelihood}

In our analyses using a CMB likelihood, we use the Planck 2018 temperature and polarization likelihoods via Plik~\cite{Planck_2018_likelihoods}. Specifically, we include the high-$\ell$ Plik TT, TE, and EE likelihood, the low-$\ell$ Commander TT likelihood, and the low-$\ell$ SimAll EE likelihood.  We neglect ACT and SPT data in this analysis.  ACT primary anisotropies and DESI DR2 show a mild tension in $\Omega_m$ when $N_{\rm{eff}}$ is varied~\cite{Camphuis_2026,sharma2026recoupleddarkradiationreconciling}.  No full ACT covariance with Planck is available and so a multipole $\ell$ cutoff is used instead to combine these datasets, but inferred cosmological parameters can shift depending on the location of a cut (demonstrated in the context of Planck PR4 in~\cite{garciaquintero2025cosmologicalimplicationsdesidr2}).  Ultimately we eschew both ACT and SPT data to avoid undue complication; all of the effects we wish to illustrate are demonstrable with Planck data alone.

Theoretical lensed CMB power spectra are computed using CLASS with lensing enabled and multipoles evaluated up to the maximum $\ell$ required by the likelihood. The lensed spectra $C_\ell^{TT}$, $C_\ell^{TE}$, and $C_\ell^{EE}$ are computed and converted to $\mu\mathrm{K}^2$ units using $T_{\mathrm{CMB}} = 2.7255\,\mathrm{K}$~\cite{Fixsen_2009} before being passed to the Planck likelihood routines.

The main cosmological parameters entering the CMB likelihood include the standard $\Lambda$CDM parameters $\{ \omega_b, \Omega_m, h, A_s, n_s, \tau_{\mathrm{reio}} \}$, as well as $N_{\rm{eff}}$ in analyses where it is not held fixed. The helium-4 mass fraction $\textrm{Y}_{\textrm{P}}=\frac{\rho_{\rm{He}}}{\rho_{\rm{H}}+\rho_{\rm{He}}}$, for helium energy density $\rho_{\rm{He}}$ and hydrogen energy density $\rho_{\rm{H}}$, is set to the value predicted by LINX for the sampled parameters, ensuring consistency between BBN and the CMB sectors.

Foreground and instrumental nuisance parameters follow the Planck recommended priors in ref.~\cite{Planck_2018_likelihoods}, or ref.~\cite{Planck2013} if priors from ref.~\cite{Planck_2018_likelihoods} are unavailable.  To avoid unphysical parameter values or numerical instabilities in CLASS, these Gaussian priors are implemented as truncated normal distributions with bounds chosen to match these recommended priors.

The full CMB log-likelihood is finally computed as the sum of the high-$\ell$ (Plik) and low-$\ell$ (Commander and SimAll) contributions:
\begin{equation}
\ln \mathcal{L}_{\mathrm{CMB}} =
\ln \mathcal{L}_{\mathrm{Plik}} +
\ln \mathcal{L}_{\mathrm{Commander}} +
\ln \mathcal{L}_{\mathrm{SimAll}}.
\end{equation}

\subsection{$N_{\rm{eff}}$}
$N_{\rm eff}$ is a measure of the relativistic degrees of freedom present in the early universe aside from photons. 
It is defined as 
\begin{equation}
    N_{\rm{eff}}\equiv\left(\frac{\rho_R-\rho_{\gamma}}{\rho_{\nu,{\rm std}}}\right)_0,
\end{equation}
where $\rho_R$ is the total energy density in radiation, $\rho_\gamma$ is the photon energy density, and $\rho_{\nu,{\rm std}}$ is the energy density of one neutrino species under the assumption of instantaneous decoupling well before $e^+e^-$ annihilation. The subscript ``0'' indicates that a quantity should be evaluated after relevant decoupling or nonrelativistic thresholds have been crossed, e.g.\ after $e^+e^-$ annihilation and neutrino decoupling in standard cosmology.
Including corrections like noninstantaneous decoupling and the finite mass of electrons leads to a precise value of $N_{\rm eff} = 3.044$ in $\Lambda$CDM~\cite{Akita:2020szl,Froustey:2020mcq,Bennett:2020zkv}. 
In analyses where $N_{\rm{eff}}$ is allowed to vary, we parametrize departures from $N_{\rm{eff}}=3.044$ as a (possibly negative) contribution to $\rho_R$ at early times, which simply redshifts as $a^{-4}$ and does not couple to Standard Model species.  This is in contrast to modifying the relative photon and neutrino temperatures to produce equivalent changes in $N_{\rm{eff}}$.

BAO and the CMB are not generally sensitive to the mechanism by which $N_{\rm{eff}}$ is modified.  BBN, however, is independently sensitive to the relative photon and neutrino temperatures and changes to the expansion rate wrought by inert radiation.  Results from varying $N_{\rm{eff}}$ are therefore model-dependent; we elect to add inert species at early times to modify $\rho_R$ as this mechanism is the most general and most consistent with the spirit of $\Lambda$CDM+$N_{\rm{eff}}$ analyses performed with CMB data alone.\footnote{e.g.\  most analyses of $\Lambda$CDM+$N_{\rm{eff}}$ using CLASS modify the number of relativistic species at early times \texttt{N\_ur}, as opposed to the ``non-cold dark matter'' temperature \texttt{T\_ncdm}, even though either produces a change in $N_{\rm{eff}}$.}

This does imply that samples with $N_{\rm{eff}}<3.044$ in our analyses do not correspond to an obvious physical scenario (see e.g.\ ref.~\cite{ganguly2026consistentnrmefffitting} for a recent discussion).  Like recent analyses suggesting an unphysical negative neutrino mass (e.g.\ refs.~\cite{Adame_2025,craig2024nusgoodnews}), preference for $N_{\rm{eff}}<3.044$ in the results reported below should be taken as an indication of a missing model component, an untracked systematic, or another consequence of mismodeling.  Further investigation would require selection of a model that injects entropy to photons, siphons entropy from neutrinos, has a low reheat temperature, or achieves a low $N_{\rm{eff}}$ through some other exotic effect (see e.g.\ refs.~\cite{Serpico_2004b,Cadamuro_2012,Berezhiani_2013,Ho_2013,Nollett_2014,de_Salas_2015,Millea_2015,Berlin_2019,Hasegawa_2019,Depta_2020,Sabti_2020,Li_2020,Giovanetti_2022,Akita_2025a,Akita_2025b,Giovanetti_2025,Barbieri_2025,Escudero_Abenza_2026,escudero2026doesnrmeff,Jung_2026,ganguly2026consistentnrmefffitting} for examples of new physics analyses that model these classes of effects directly).
We stress that in such scenarios, there are no available summaries that can be used in place of the full BBN likelihood used here. 

\section{Results}
Using the likelihoods and methods described above, we perform analyses on combinations of CMB, BAO, and BBN data: CMB alone, BBN alone, BAO+CMB, BAO+BBN, BBN+CMB, and BAO+BBN+CMB.  We do not discuss results of the CMB, BBN, BAO+CMB, and BBN+CMB analyses at length, as each has already been performed in the literature (e.g.\ refs.~\cite{Planck2018,calabrese2025atacamacosmologytelescopedr6,Camphuis_2026,Pitrou_2021,Pisanti_2021,Fields_2020,Yeh_2021,Adame_2025,Schoneberg_2019,schoneberg_2024,LINX_short,launders2026,goldstein20262determinationnrmeff}) and our results are largely consistent modulo predictable changes from different reaction networks, different dataset selection, and an updated helium-4 abundance.  Our full results are included in appendix~\ref{app:results} for comparison; for BBN alone, BAO+BBN, BBN+CMB, and BAO+BBN+CMB, we obtain and report results in this appendix in $\Lambda$CDM (with $N_{\rm{eff}}=3.044$), and $\Lambda$CDM+$N_{\rm{eff}}$; for CMB and BAO+CMB, since they have been explored extensively and are not the focus of this manuscript, we report only $\Lambda$CDM+$N_{\rm{eff}}$.

Measurements of primordial helium-4 are especially relevant when $N_{\rm{eff}}$ is allowed to vary.  Recent results from refs.~\cite{aver_2026,yeh2026} have placed tight constraints on the abundance of primordial helium-4, and correspondingly tight constraints on $N_{\rm{eff}}$.  With this new measurement, BBN now provides the most stringent constraints on $N_{\rm{eff}}$ out of any individual cosmological epoch.

\subsection{BAO+BBN}\label{sec:BAO_BBN_no_Neff}

\begin{figure}
\centering
\includegraphics[width=.49\linewidth]{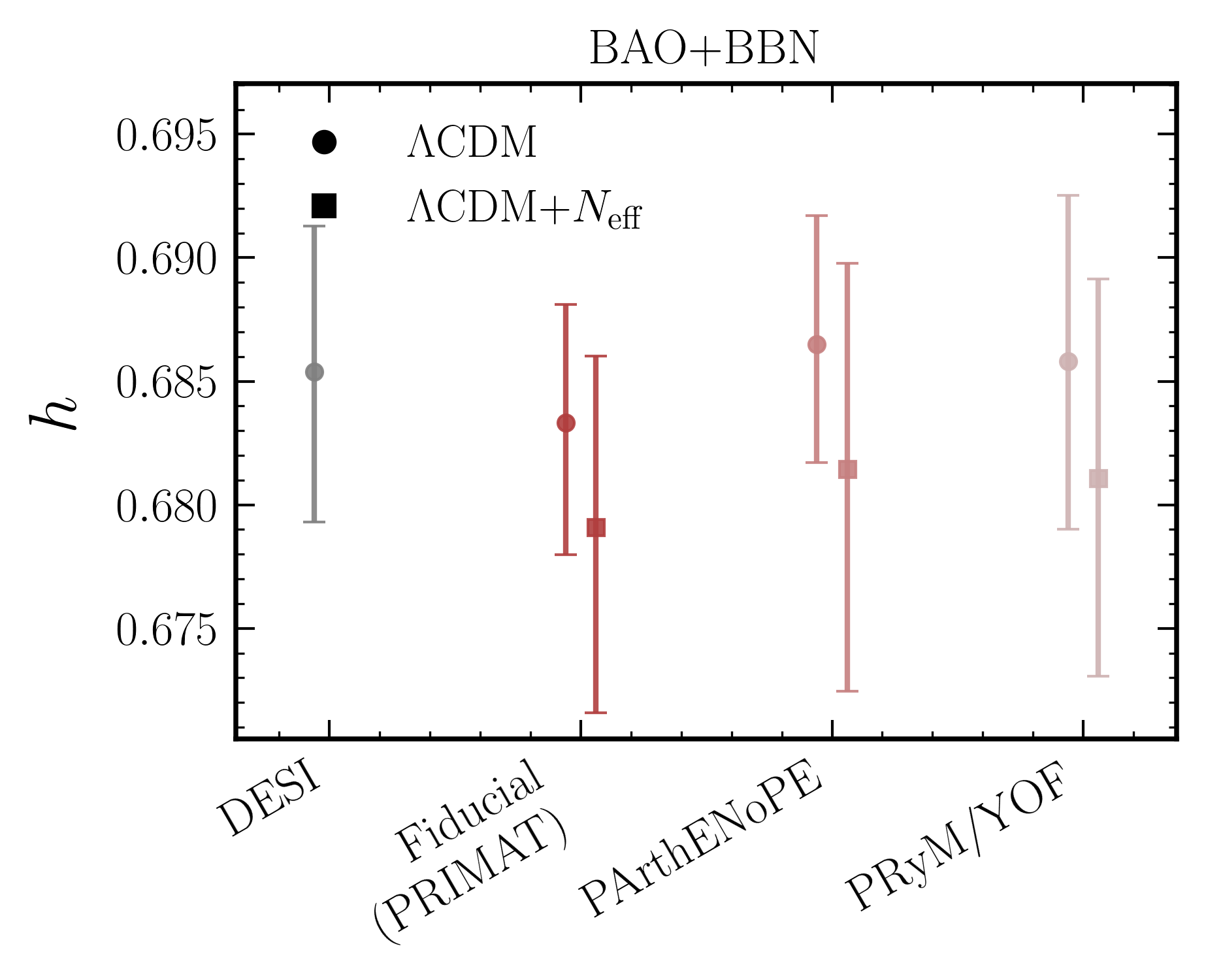}
\includegraphics[width=0.49\linewidth]{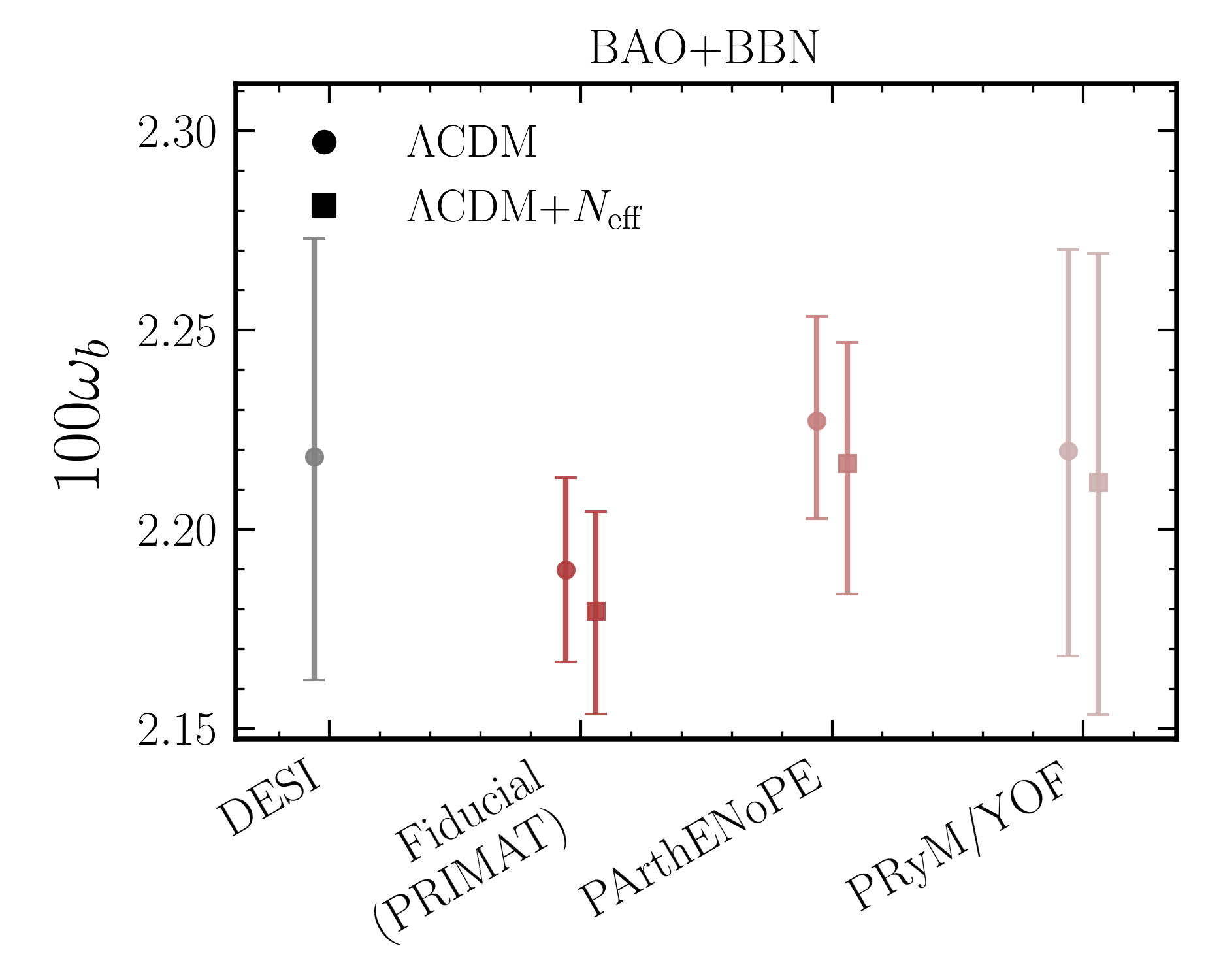}
\caption{\textit{Left:} Medians and $\pm 1\sigma$ uncertainties for the dimensionless equivalent of the Hubble constant $h$, for the three different reaction networks considered in this work in a joint BAO+BBN analysis, and the value obtained in the DESI DR2 BAO analysis.  Note that DESI uses the PRyM/YOF network.  Circular markers are used for $\Lambda$CDM results, with square markers for $\Lambda$CDM+$N_{\rm{eff}}$.  \textit{Right}: Medians and $\pm 1\sigma$ uncertainties for the baryon density from our BAO+BBN analysis for all three reaction networks, compared to the DESI DR2 result.  Since BAO provides little information about the baryon density, the reaction network entirely determines $\omega_b$, presenting a significant systematic for BAO+BBN analyses.}
\label{fig:BAO_BBN}
\end{figure}

We perform an analysis of DESI DR2 BAO and BBN data, to obtain a CMB-independent inference of $h$.  We summarize these results and compare with results reported by the DESI collaboration~\cite{DESI:2025zgx} in figure~\ref{fig:BAO_BBN}.  The DESI result uses the PRyM/YOF network to place a prior on the baryon density; by contrast, our result uses the full BBN likelihood described in section~\ref{sec:bbn_like}.  Our fiducial result (using the PRIMAT reaction network) is 
\begin{equation}
    h=0.6833_{-0.0053}^{+0.0048}\qquad \textrm{Fiducial (PRIMAT), BAO+BBN}.\label{eq:h_BAO_BBN}
\end{equation}
We find a modest spread in $h$ across the three reaction networks considered here, up to roughly $0.5\sigma$.  While this difference is not significant, it illustrates that Hubble tension analyses that use the PRIMAT reaction network or a derivative---our recommendation---must resolve a tension of even larger significance than those using the PArthENoPE or PRyM/YOF networks, even without CMB information.

We also show inferred baryon density in figure~\ref{fig:BAO_BBN}.  Since BAO depends on the BBN determination of baryon density to break parameter degeneracies, the choice of reaction network is a large systematic in this analysis.  These results are consistent with ref.~\cite{LINX_short}. 

We also perform a BAO+BBN analysis in which $N_{\rm{eff}}$ is allowed to float.  As far as the authors are aware, this is the first time such an analysis has been performed in the literature with DESI DR2, and with a full BBN likelihood marginalizing over nuclear rates, providing a CMB-independent measurement of $h$, $\omega_b$, and $N_{\rm{eff}}$.  These results for $h$ and $\omega_b$ in $\Lambda$CDM+$N_{\rm{eff}}$ are also shown in figure~\ref{fig:BAO_BBN}.

We find a smaller $h$ is preferred by the combination of BAO and BBN when $N_{\rm{eff}}$ is allowed to float.  In our fiducial analysis, 
\begin{align}
    h&=0.6791^{+0.0069}_{-0.0075}\nonumber\\
    N_{\rm{eff}}&=2.974^{+0.098}_{-0.099}\qquad \textrm{Fiducial (PRIMAT), BAO+BBN}.\label{eq:h_Neff_BAO_BBN}
\end{align}
This reduced $h$ is consistent with a preference for a slightly smaller central value of $N_{\rm{eff}}$ than in $\Lambda$CDM, leading to slightly larger $r_d$ (with $hr_d$ being the parameter combination that DESI is most sensitive to). 

\subsection{BAO+BBN+CMB}\label{sec:all3}

\begin{figure}
    \centering
    \includegraphics[width=0.7\linewidth]{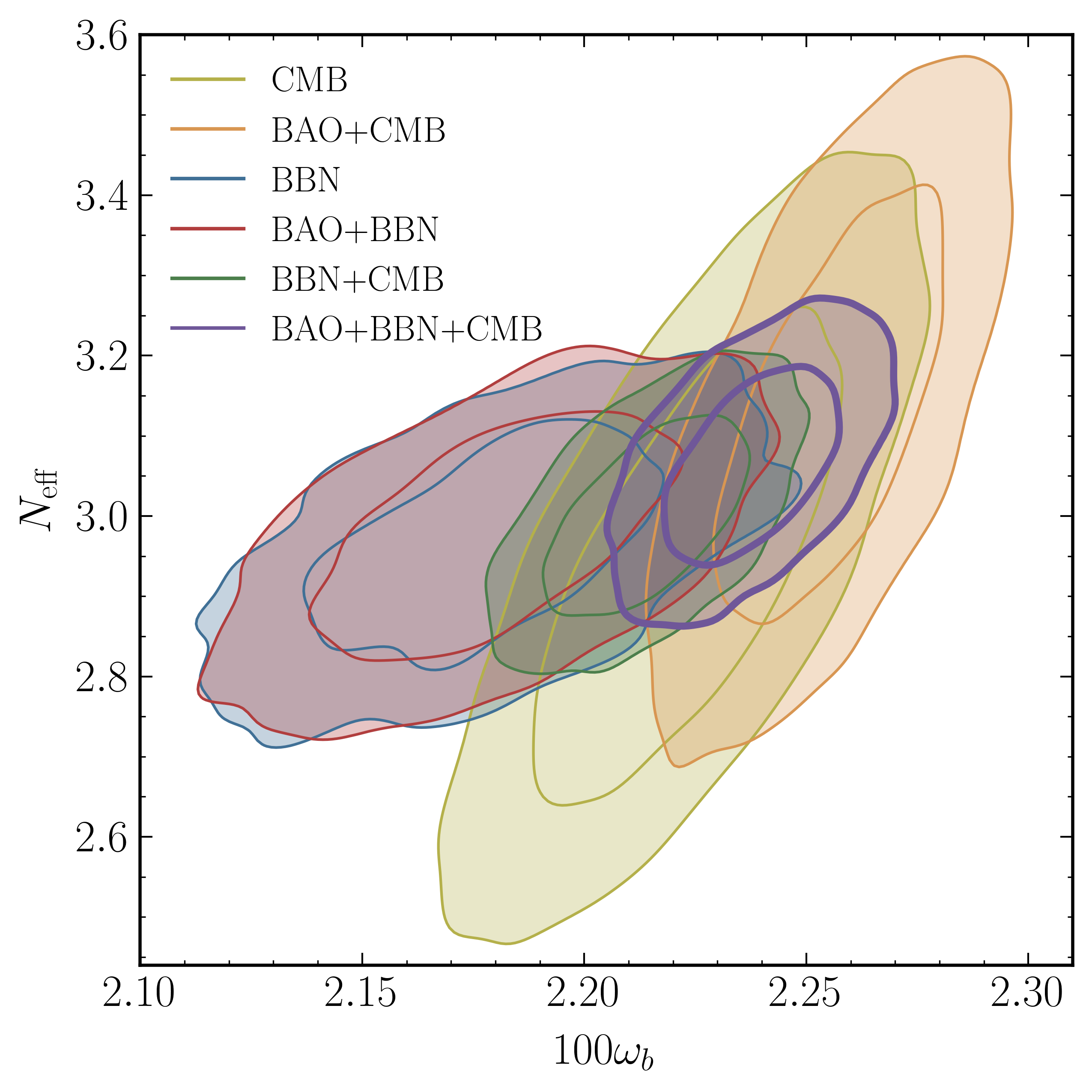}
    \caption{Fiducial results in the $N_{\rm{eff}}-\omega_b$ plane for all dataset combinations considered in this work, for $\Lambda$CDM+$N_{\rm{eff}}$.  The result from the combined BAO+BBN+CMB analysis is shown in purple. }
    \label{fig:all}
\end{figure}

With our BAO+BBN results reported, we now move on to a joint combination with the CMB.  As discussed in section~\ref{sec:intro}, this data combination has been overlooked by major collaborations, and is being investigated here with a full likelihood for the first time in the literature.  Our results from parameter estimation are summarized in appendix~\ref{app:results}, and results from the $\Lambda$CDM+$N_{\rm{eff}}$ analysis are illustrated in figure~\ref{fig:all}. In what follows, we highlight nontrivial parameter behavior in this data combination.

\begin{figure}
    \centering
    \includegraphics[width=0.7\linewidth]{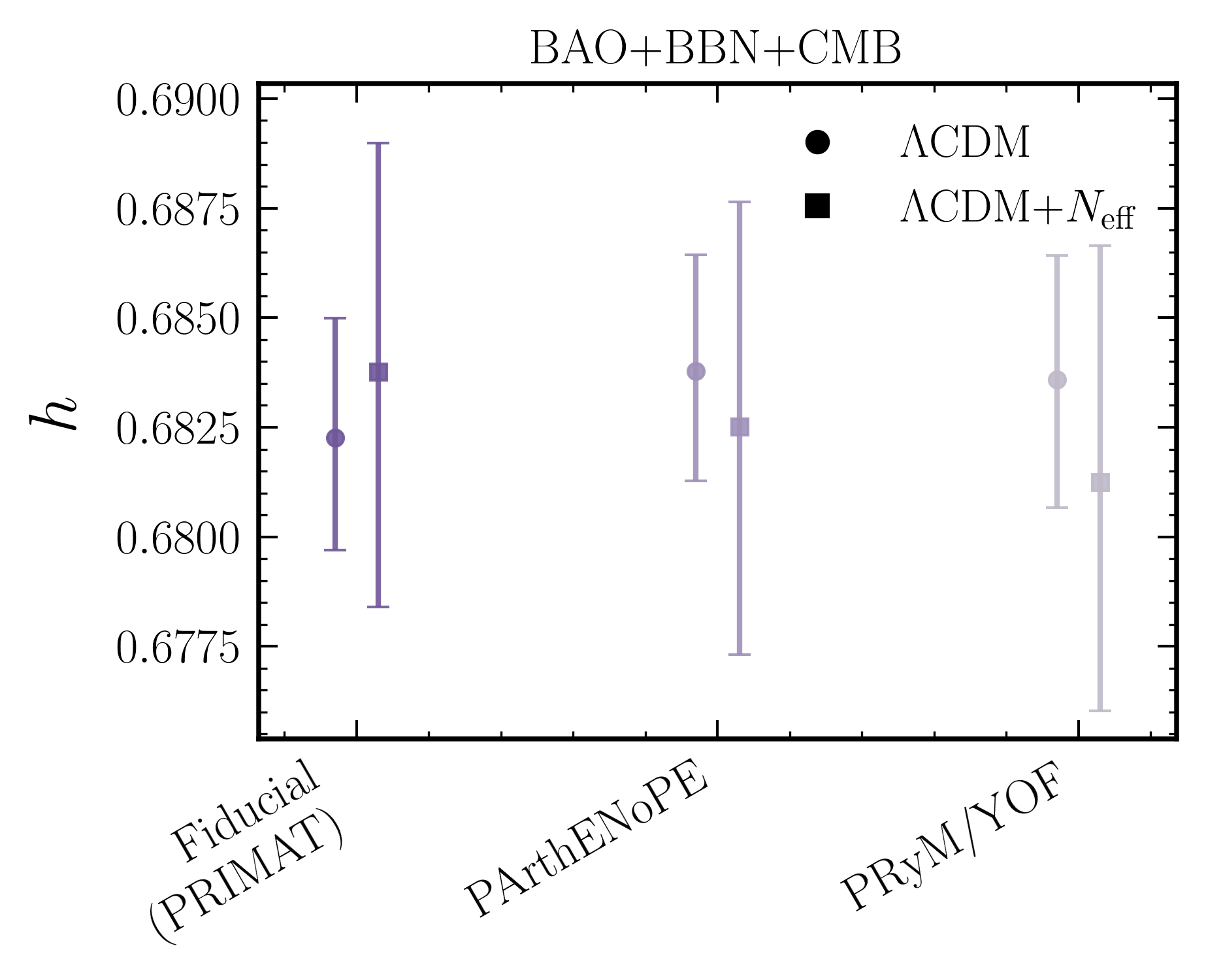}
    \caption{Determination of $h$ for three different reaction networks, in $\Lambda$CDM (circular markers) and $\Lambda$CDM+$N_{\rm{eff}}$ (square markers).  While the PRIMAT network favors the lowest $h$ in $\Lambda$CDM, it prefers the highest $h$ of the three networks when $N_{\rm{eff}}$ is allowed to float.}
    \label{fig:h_all}
\end{figure}

We summarize our determination of $h$ in figure~\ref{fig:h_all}.  When $N_{\rm{eff}}$ is fixed, we find the following values for $h$ with this combination of datasets:
\begin{alignat}{2}
h&=0.6823^{+0.0027}_{-0.0026}\qquad &&\textrm{Fiducial (PRIMAT), BAO+BBN+CMB}\nonumber\\
h&=0.6838^{+0.0027}_{-0.0025}\qquad &&\textrm{PArthENoPE, BAO+BBN+CMB}\nonumber\\
h&=0.6836^{+0.0028}_{-0.0029}\qquad &&\textrm{PRyM/YOF, BAO+BBN+CMB},\label{eq:h_all}
\end{alignat}
which, as in the BAO+BBN results in section~\ref{sec:BAO_BBN_no_Neff}, show a spread of roughly $0.5\sigma$.  These results are consistent with expectation: of the three reaction networks, the PRIMAT reaction network prefers the lowest $\omega_b$, which has a positive correlation with $h$ in the CMB.  

However, the results change qualitatively when $N_{\rm{eff}}$ is allowed to float, and we find in this scenario that the PRIMAT network gives the largest $h$:
\begin{alignat}{2}
h&=0.6838^{+0.0052}_{-0.0054},\,N_{\rm{eff}}=3.065^{+0.078}_{-0.076}\qquad &&\textrm{Fiducial (PRIMAT), BAO+BBN+CMB}\nonumber\\
h&=0.6825^{+0.0051}_{-0.0052},\,N_{\rm{eff}}=3.027\pm0.079\qquad &&\textrm{PArthENoPE, BAO+BBN+CMB}\nonumber\\
h&=0.6812^{+0.0054}_{-0.0052},\,N_{\rm{eff}}=2.991^{+0.081}_{-0.080}\qquad &&\textrm{PRyM/YOF, BAO+BBN+CMB}.\label{eq:h_neff_full}
\end{alignat}
This behavior arises because DESI DR2 prefers a low $\Omega_m$ as compared to Planck CMB~\cite{DESI:2025zgx}.  To compensate, a combination of CMB and BAO data prefers a larger $\omega_b$, $h$, and $N_{\rm{eff}}$ to mitigate a lower $\Omega_m$, compared to just the CMB; this can be seen by comparing the CMB and BAO+CMB contours in figure~\ref{fig:all}.  On the other hand, our fiducial PRIMAT BBN result also informs the value of $\omega_b$ and $N_{\rm{eff}}$.  The combination of BBN with BAO+CMB ultimately prefers a large $N_{\rm{eff}}$, which in turn drives the preference for a larger value of $h$ in spite of a slightly lower $\omega_b$.

\subsubsection{BBN nuisance parameters}
\begin{figure}
    \centering
    \includegraphics[width=0.45\linewidth]{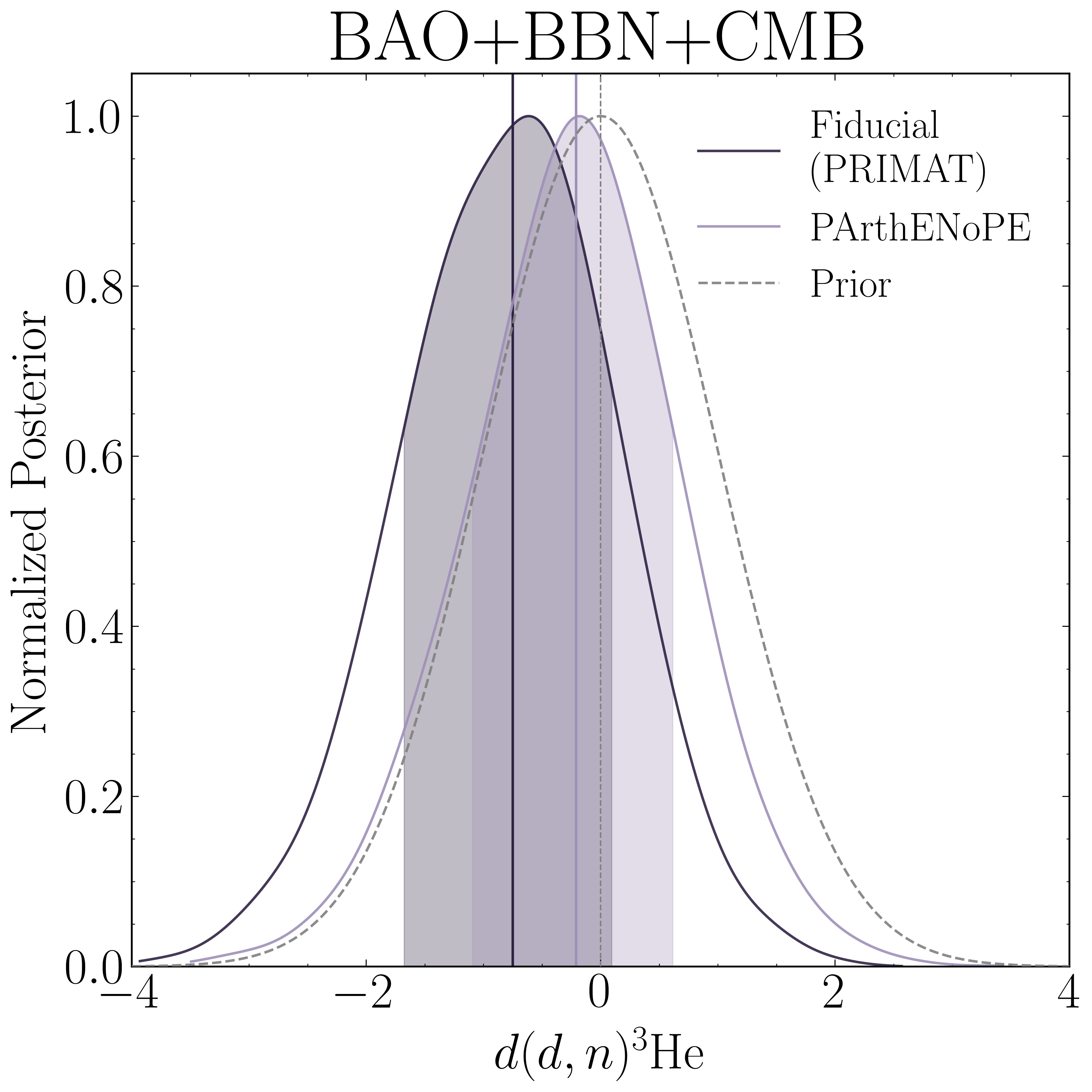}
    \includegraphics[width=0.45\linewidth]{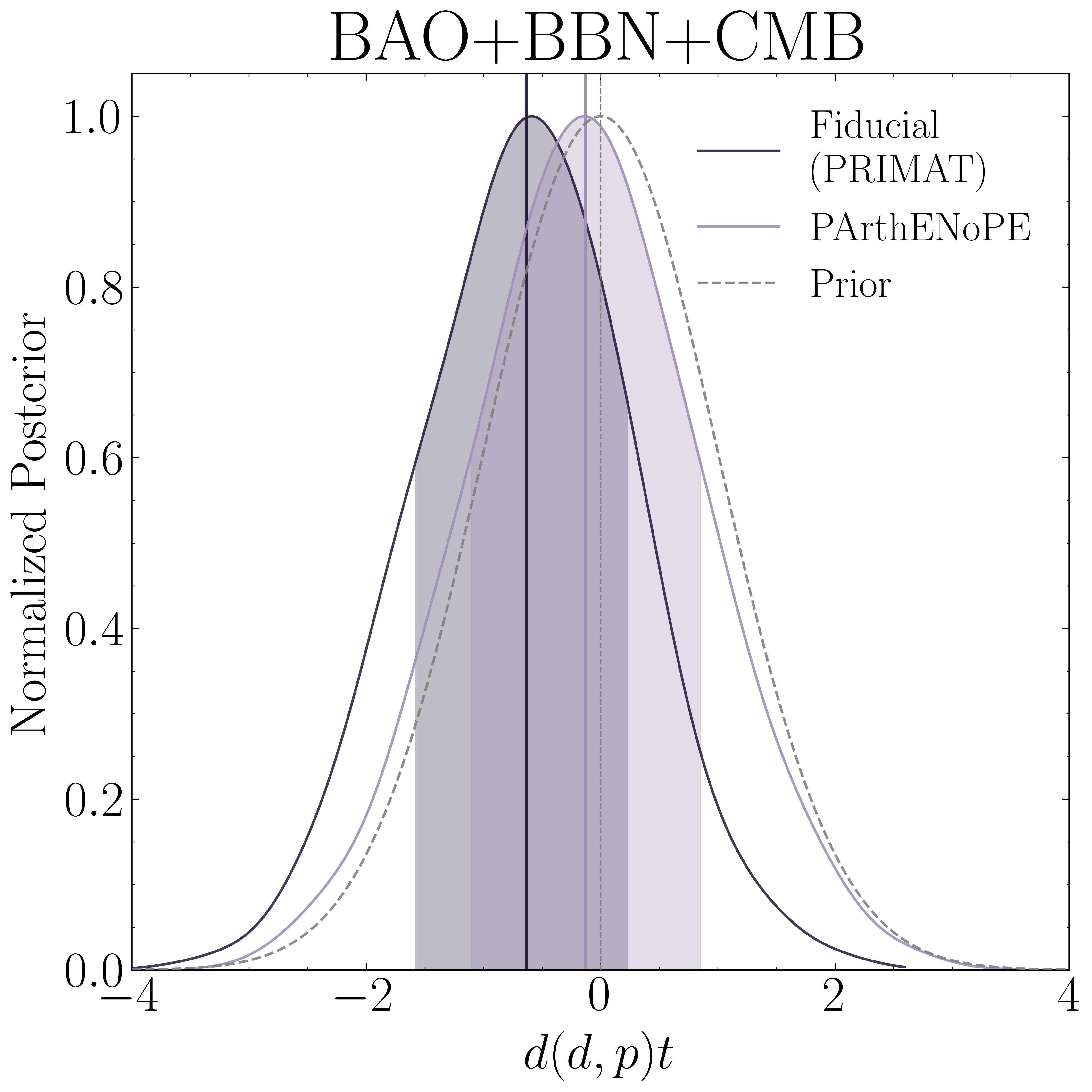}
    \includegraphics[width=0.45\linewidth]{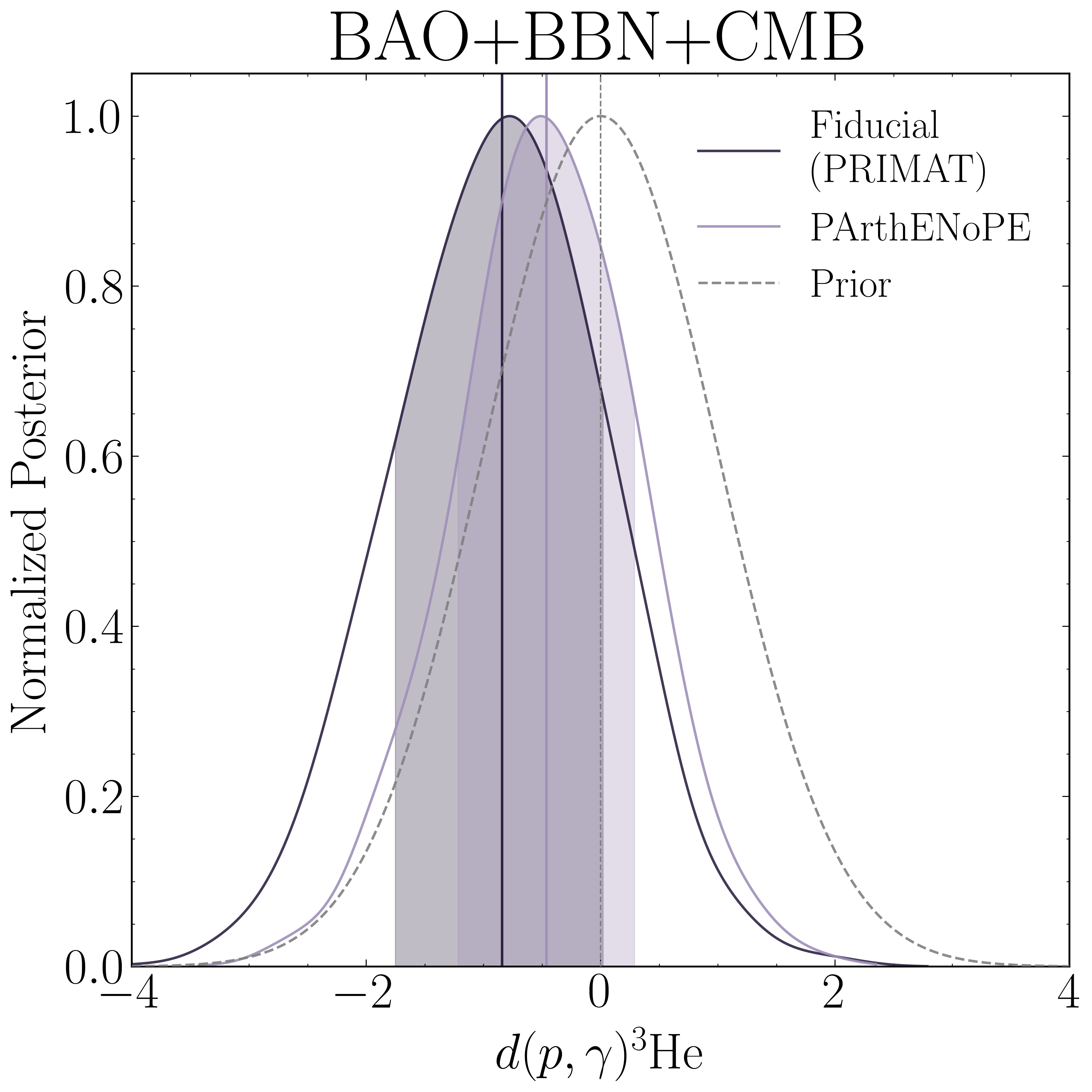}
    \caption{Posteriors for three BBN nuisance parameters describing the rates of the three reactions most critical for determining the primordial deuterium abundance, in the joint BAO+BBN+CMB analysis.  Dashed gray distributions are the $\mathcal{N}(0,1)$ priors from which these parameters are drawn.  The posteriors using both the PRIMAT and PArthENoPE reaction networks are shown.  Note that even the PArthENoPE posterior for $d(p,\gamma)^3$He is shifted noticeably from the prior at $\mathcal{N}(0,1)$ under this data combination.}
    \label{fig:dpg}
\end{figure}

One advantage of using LINX for BBN is that it allows the user to simultaneously vary BBN nuisance parameters controlling nuclear rate uncertainties, and obtain posteriors in those nuisance parameters.  Shifts in nuisance parameter posteriors were noted in ref.~\cite{LINX_short} when the PRIMAT BBN predictions were combined with CMB data.  Here, we note that those shifts become even more pronounced when BAO data are included.  Figure~\ref{fig:dpg} shows the 1D posteriors for three such nuisance parameters, most important for determining the primordial deuterium abundance.  For the reaction $d(p,\gamma)^3\rm{He}$, the posterior is shifted by nearly one sigma when the PRIMAT reaction network is used, and $>0.5\sigma$ when the PArthENoPE network is used.

While these shifts are minor, we highlight this particular shift since $d(p,\gamma)^3\rm{He}$ is the best-measured of the three nuclear rates with the strongest impact on the deuterium abundance.  The PRIMAT and PArthENoPE reaction networks are in agreement about the appropriate interpolation to use for this rate~\cite{Pitrou_2021,Pisanti_2021}; in other words, there is little to no reaction network systematic for this rate.  Yet, for concordance with the CMB and BAO, this rate must shift; when the PRIMAT rates are used for $d(d,p)^3\rm{H}$ and $d(d,n)^3\rm{He}$, the $d(p,\gamma)^3\rm{He}$ rate must shift more significantly than the others, and even when the PArthENoPE rates are used there is some small preference to shift $d(p,\gamma)^3\rm{He}$ away from its prior mean. 

This behavior is due to the elevated sensitivity of the deuterium abundance to this particular reaction rate compared to its uncertainty~\cite{LINXlong}; a large number density of protons makes this rate particularly impactful for deuterium.  The noticeable shift in this rate arises from differences between the BAO+CMB-preferred $\omega_b$ and the $\omega_b$ inferred from the deuterium prediction \textit{in either network}; even the inference for $\omega_b$ with PArthENoPE is slightly low compared to the BAO+CMB result (see figure~\ref{fig:all}).  This suggests more precise measurements of $d(p,\gamma)^3\rm{He}$ will continue to have an outsize effect on the precision of joint cosmological predictions, despite the relative precision with which it is already measured compared to $d(d,p)^3\rm{H}$ and $d(d,n)^3\rm{He}$.  These results are also indicative of the relative precision between the observed D/H and the nuclear experimental data feeding into the prediction of D/H; the astrophysical data are currently \textit{more} precise, so much so that the rate nuisance parameters can be inferred from astrophysics at higher precision than the direct measurements of these rates, assuming $\Lambda$CDM.  This is true even in the absence of BAO or CMB data, though figure~\ref{fig:dpg} and comparison with ref.~\cite{LINX_short} indicate that additional information about the relevant nuclear physics may also be available from these other probes.

\subsubsection{$N_{\rm{eff}}$ precision and network dependence}\label{sec:Neff_from_DH}
\begin{figure}
    \centering
    \includegraphics[width=0.49\linewidth]{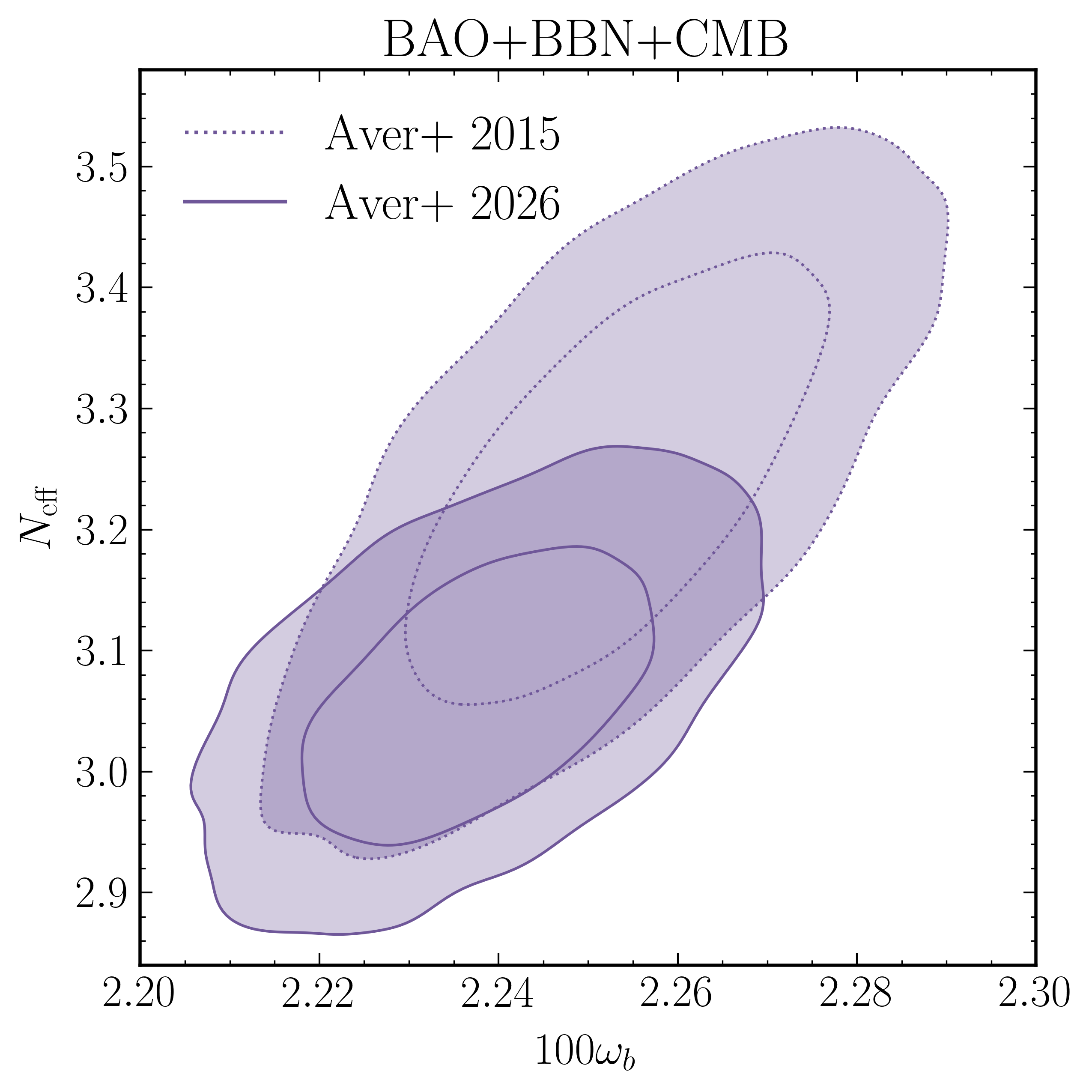}
    \includegraphics[width=0.49\linewidth]{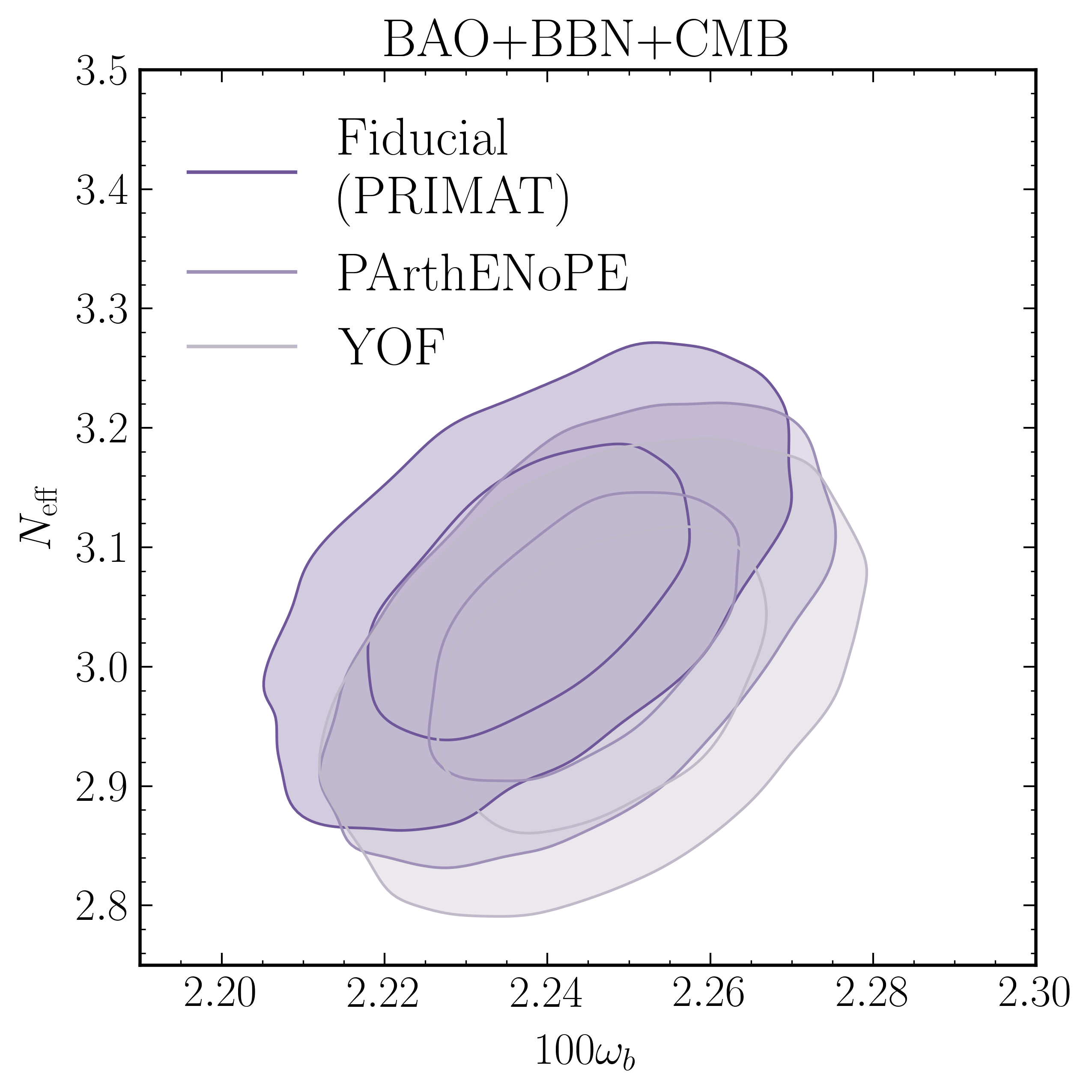}
    \caption{Results in the $100\omega_b-N_{\rm{eff}}$ plane for our fiducial BAO+BBN+CMB analysis.  \textit{Left:}  Fiducial results using Y$_{\rm{P}}$ from ref.~\cite{aver_2026} (``Aver+ 2026'') and ref.~\cite{Aver_2015} (``Aver+ 2015'').  As expected, given the sensitivity of Y$_{\rm{P}}$ to $N_{\rm{eff}}$ and the tighter observational constraints on Y$_{\rm{P}}$ from ref.~\cite{aver_2026}, the full joint analysis returns much tighter constraints on $N_{\rm{eff}}$ than could be obtained with older Y$_{\rm{P}}$ measurements.  \textit{Right: } Results using Y$_{\rm{P}}$ from ref.~\cite{aver_2026} for all three reaction networks.}
    \label{fig:full_old_new}
\end{figure}

To explore our results in $N_{\rm{eff}}$, we begin by highlighting the effect of new helium-4 observations from ref.~\cite{aver_2026} on this joint analysis in figure~\ref{fig:full_old_new}.  The contours shrink considerably in the $N_{\rm{eff}}$ direction, as expected, when the new measurement is used.

This reduction in uncertainty, combined with the insensitivity of the predicted Y$_{\rm{P}}$ to reaction network, may lead the reader to conclude that $N_{\rm{eff}}$ is driven primarily by the BBN determination of Y$_{\rm{P}}$.  The resulting $N_{\rm{eff}}$ prediction would therefore be insensitive to choice of reaction network or the primordial deuterium abundance, as this systematic primarily impacts D/H.  When BBN is used on its own to predict $N_{\rm{eff}}$, these are reasonable conclusions to draw.

However, when combined with the CMB, this is no longer a valid inference.  $N_{\rm{eff}}$ in our BAO+BBN+CMB combinations is shown in eq.~\eqref{eq:h_neff_full}, and show a spread of roughly $1\sigma$.  This result is illustrated in the right panel of figure~\ref{fig:full_old_new}.

\begin{figure}
    \centering
    \includegraphics[width=0.75\linewidth]{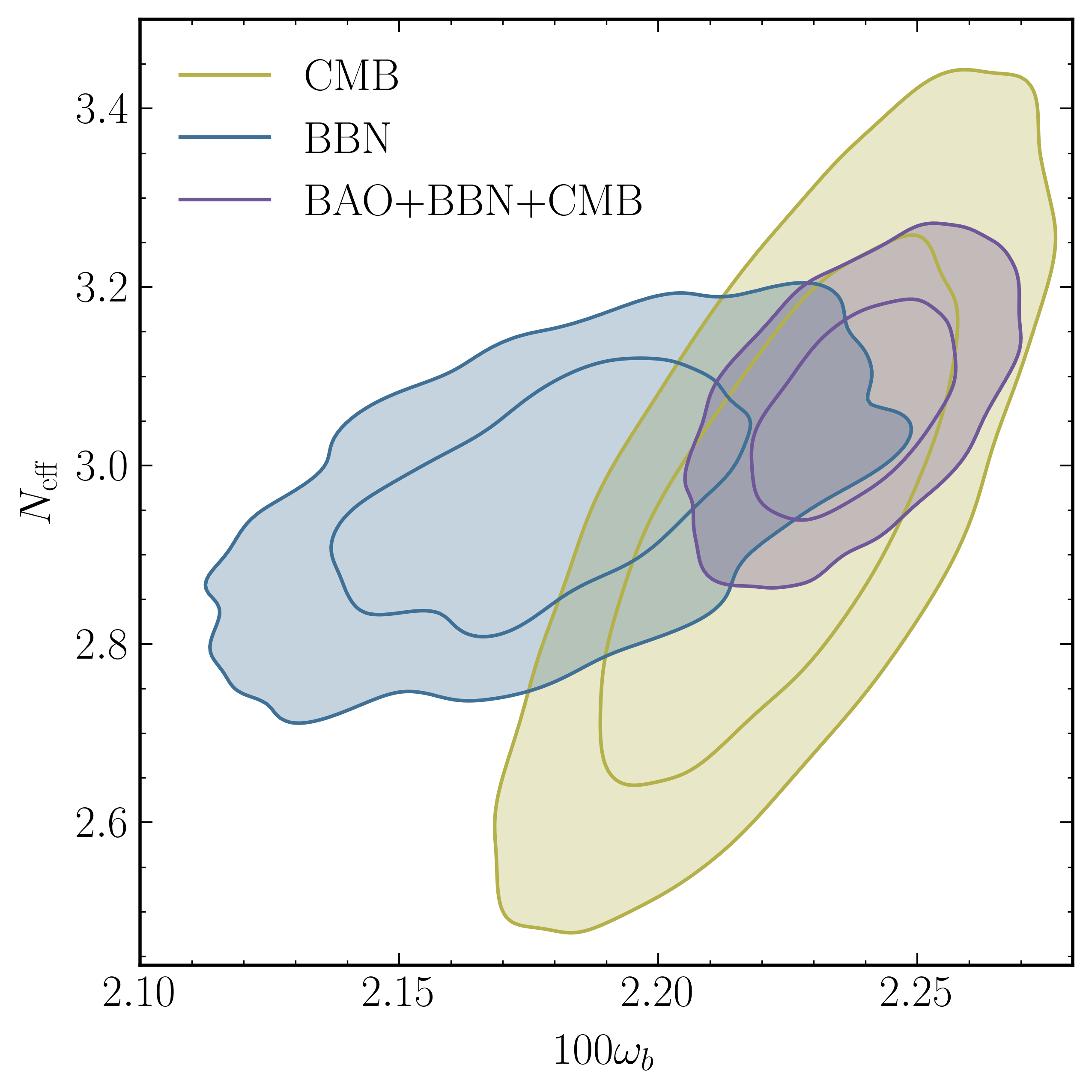}
    \caption{Results in the $100\omega_b-N_{\rm{eff}}$ plane, using the CMB-only likelihood (beige green) and the BBN-only likelihood (blue).  The full BAO+BBN+CMB result is also shown for reference in purple.  Because the degeneracy directions between these two parameters are different in these two independent epochs, and because there is mild disagreement in the preferred $\omega_b$ across these epochs, the contours overlap at higher $N_{\rm{eff}}$ than is preferred by either epoch independently.}
    \label{fig:BBN_CMB}
\end{figure}

This dependence is due entirely to the spread in the predicted mean and uncertainty of the deuterium abundance with different choices of reaction network.  As noted in ref.~\cite{LINX_short}, the different degeneracy directions between $N_{\rm{eff}}$ and $\omega_b$ are not entirely aligned in BBN and the CMB.  Since the PRIMAT network prefers a low value of $\omega_b$ as compared to the CMB, the highest likelihood region is where the two contours intersect, at higher $N_{\rm{eff}}$ than preferred by either the CMB or BBN independently.  This behavior is highlighted in figure~\ref{fig:BBN_CMB} using the PRIMAT network.  The other two networks recover lower $N_{\rm{eff}}$, given the relative agreement between the $\omega_b$ preferred by the PArthENoPE network and the CMB, and given the large error bars on the determination of $\omega_b$ using the PRyM/YOF network.
The end result is that the three networks have a roughly $1\sigma$ discrepancy in their inferred values of $N_{\rm eff}$ with all datasets combined. 

The spread in the inferred $N_{\rm{eff}}$ is important to recognize, because it has implications for which distinct classes of models are favored or disfavored by data.  $N_{\rm{eff}}=3.044$, being the Standard Model prediction for $N_{\rm{eff}}$~\cite{Akita:2020szl,Froustey:2020mcq,Bennett:2020zkv}, represents a bifurcation in phenomenological models, where the mechanisms required to achieve $N_{\rm{eff}}>3.044$ are largely orthogonal to those required for $N_{\rm{eff}}<3.044$ (see ref.~\cite{escudero2026doesnrmeff} for a recent discussion in light of the ACT preference for low $N_{\rm{eff}}$).  While all of our results are consistent with $N_{\rm{eff}}=3.044$---i.e.\  no new physics is required for consistency with data---whether $N_{\rm{eff}}$ tolerates positive or negative departures from $3.044$ provides an important guardrail when building models of new physics or attempting to resolve other cosmological tensions.

\subsubsection{BBN prior versus full likelihood}

Finally, we discuss the impact of using priors on BBN predictions and uncertainties, versus constructing a full likelihood.

As discussed in section~\ref{sec:intro}, it is common for analyses involving BBN to summarize statistical uncertainties rather than marginalizing over nuisance parameters.  Since the dependence of $\omega_b$ on the reaction rates is nonlinear, it is not \textit{a priori} evident that this is a reasonable approximation.  Further, this procedure requires the existence of a suitable posterior to be used as a prior, which may not always be available, e.g.\ for models beyond $\Lambda$CDM and $\Lambda$CDM+$N_{\rm eff}$, or when nuclear reaction rates receive important updates that significantly change the posterior. 

Here we investigate the consequences of two different methods for summarizing information from BBN in a likelihood.  In the first, we simply place a Gaussian prior on $\omega_b$ inferred from D/H.  When $N_{\rm{eff}}$ is allowed to float, this procedure fails to capture the correlation between $N_{\rm{eff}}$ and $\omega_b$.  In the second summary procedure, we follow a procedure similar to that used in ref.~\cite{calabrese2025atacamacosmologytelescopedr6} and use a fixed error bar for D/H rather than varying nuisance parameters.  We estimate the uncertainty for each network by drawing 4,000 combinations of the 12 BBN nuisance parameters that scale reaction rates at $(\omega_b,N_{\rm{eff}})=(0.02237,3.044)$ to estimate a theory uncertainty $\hat\sigma_{\rm{D/H}}^{\rm{th}}$.  There is an additional, observational contribution to the uncertainty of D/H, $\sigma_{\rm{D/H}}^{\rm{obs}}$.  We add these in quadrature to construct a modified BBN likelihood
\begin{equation}
\ln\tilde{\mathcal{L}}_{\rm{BBN}} = -\frac{1}{2}\left[\left(\frac{\textrm{Y}_{\textrm{P}}^{\rm{BBN}}-\textrm{Y}_{\textrm{P}}^{\rm{BBN,obs}}}{\sigma_{\textrm{Y}_{\textrm{P}}^{\rm{BBN,obs}}}}\right)^{2}+\left(\frac{\textrm{D/H}-\textrm{D/H}^{\rm{obs}}}{\sqrt{\left(\hat\sigma_{\rm{D/H}}^{\rm{th}}\right)^2+\left(\sigma_{\textrm{D/H}^{\rm{obs}}}\right)^2}}\right)^{2}\right].\nonumber
\end{equation}
We compute D/H and Y$_{\rm{P}}$ with LINX at each sampled $(\omega_b,N_{\rm{eff}})$ pair and do not sample the nuisance parameters describing the BBN rates, as their effects have already been summarized (Y$_{\rm{P}}$ does not depend strongly on any of these rates).  We still sample the neutron lifetime alongside the model parameters to capture its effect on Y$_{\rm{P}}$.

This procedure is more principled than the first in that it can still capture the correlation between $\omega_b$ and $N_{\rm{eff}}$.  However, it neglects the increased spread in the D/H prediction in scenarios where $N_{\rm{eff}}$ is allowed to vary, which is not accounted for in analyses that utilize this summary~\cite{calabrese2025atacamacosmologytelescopedr6,goldstein20262determinationnrmeff}.  Further, given the nonlinear relationship between the BBN nuisance parameters and the prediction for D/H, potentially resulting in non-Gaussianity, this procedure still risks a poorer estimation of the posterior than the full likelihood. 

\begin{figure}
    \centering
    \includegraphics[width=0.45\linewidth]{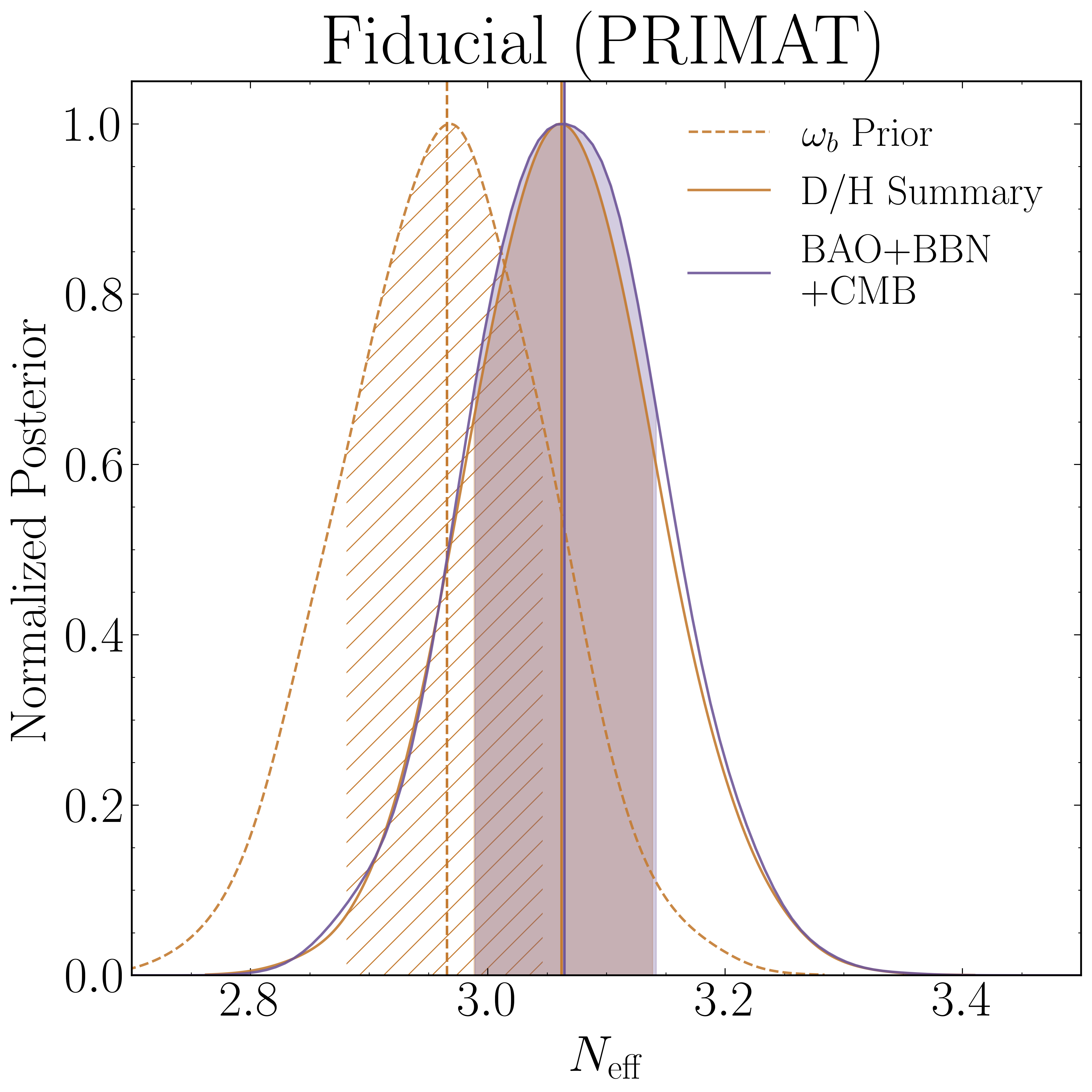}
    \includegraphics[width=0.45\linewidth]{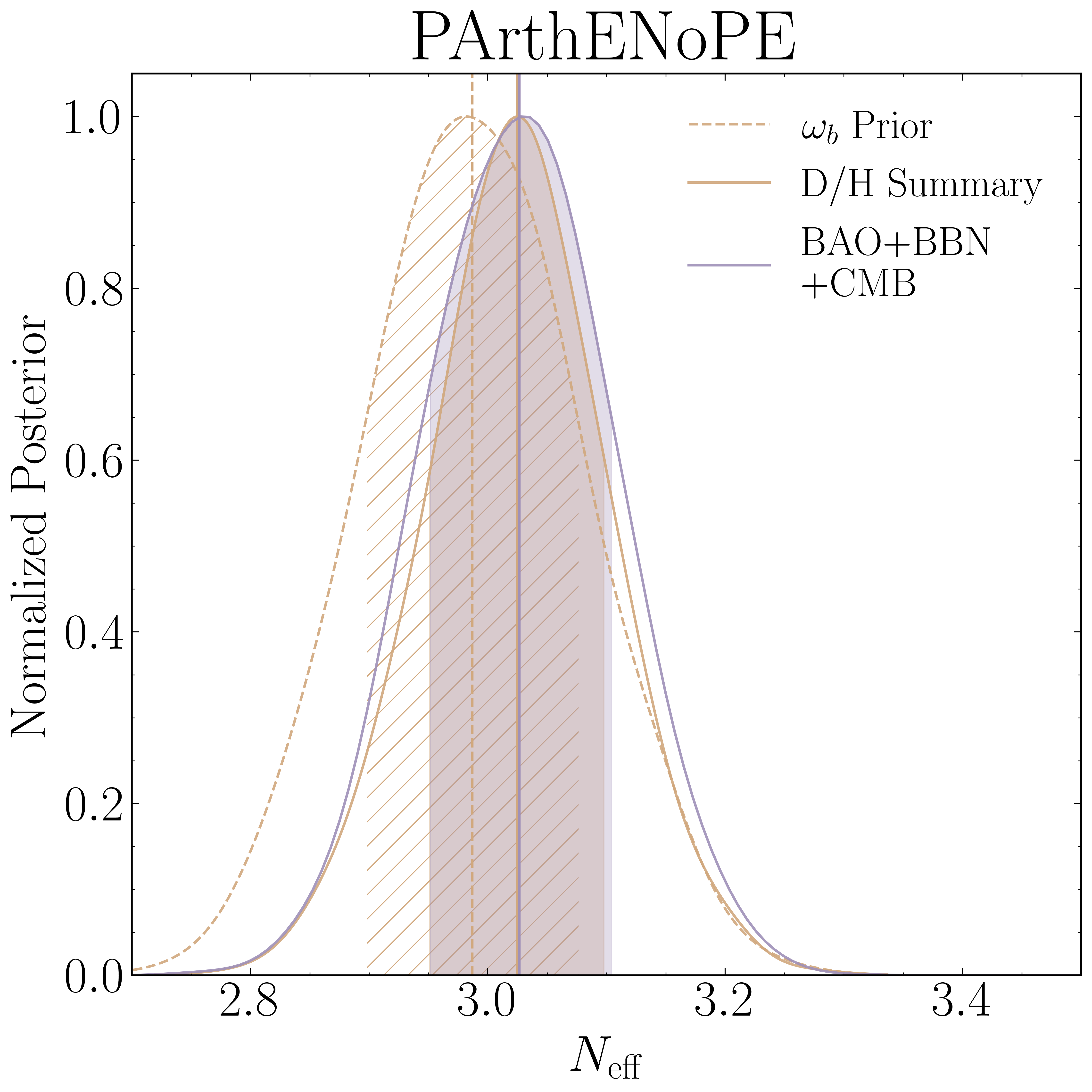}
    \includegraphics[width=0.45\linewidth]{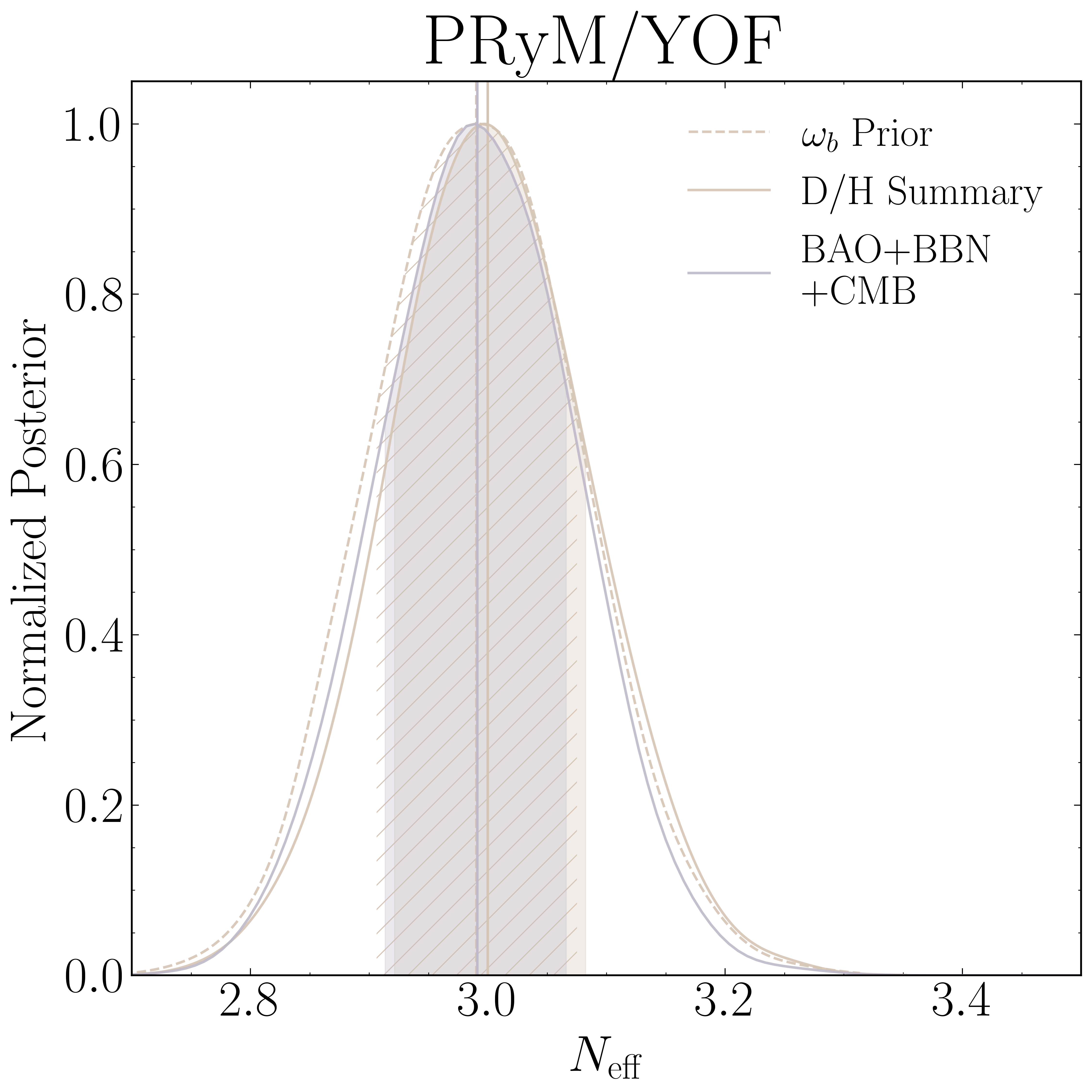}
    \caption{$N_{\rm{eff}}$ posteriors in three different BAO+BBN+CMB analyses.  In the first, the BBN likelihood is summarized through a 1D prior on $\omega_b$ (``$\omega_b$ prior'').   In the second, the BBN likelihood is summarized by a constant theory and observational uncertainty on the deuterium abundance (``D/H Summary'').  The results from the BAO+BBN+CMB analysis with a full BBN likelihood are also included for reference (``BAO+BBN+CMB'').  See discussion in text for more details about each analysis.}
    \label{fig:prior_vs_full}
\end{figure}

Results from each of these procedures in analyses where $N_{\rm{eff}}$ is allowed to float are summarized in figure~\ref{fig:prior_vs_full}.

The first procedure, which simply places a Gaussian prior on $\omega_b$, produces the wrong behavior of $N_{\rm{eff}}$ across reaction networks.  While in the full analysis the PRIMAT network produces the largest $N_{\rm{eff}}$ of the three reaction networks, here it produces the smallest of the three, entirely because the correlation between $\omega_b$ and $N_{\rm{eff}}$ that produced the high $N_{\rm{eff}}$ in the full analysis is absent from the BBN prior used here.

Despite the concerns enumerated above, we find that the second summary procedure, which pre-marginalizes the uncertainty on D/H, produces a reasonable approximation of the full posterior.  This provides critical validation of literature results that use this procedure.  However, we note that the reproduction is not exact, and in particular the difference in the upper bound on $N_{\rm{eff}}$ in the summarized and full analyses, using the PRyM/YOF network, is about 15\% of a standard deviation.  While the magnitude of this effect is small, it is similar to the increase in precision claimed in ref.~\cite{goldstein20262determinationnrmeff} when combining BBN with ACT CMB data.  That analysis used a similar summarized procedure with uncertainties on the theoretical prediction on D/H estimated using the PRyM/YOF network, though the present work does not include ACT CMB data and so a direct comparison is not possible.

Much as the Planck-lite likelihoods, with their nuisance parameters pre-marginalized, can be used in applications that do not require authoritative results or extraordinary precision, we have similarly shown that a pre-marginalized BBN likelihood provides a means to achieve a good approximation to the full posterior in $\Lambda$CDM, and $\Lambda$CDM+$N_{\rm{eff}}$, for the reaction networks considered here, using the Planck data.  In applications demanding higher precision, we recommend the use of the full likelihood, much as the full Planck likelihoods are sometimes preferred over the lite likelihoods.  We also note that additional validation of a compressed likelihood may be required for scenarios not considered here; given that the only way to check the validity of the assumptions that go into flattening the full BBN likelihood into a prior is to perform the equivalent full analysis, and given that running the full analysis is not significantly more expensive than the summarized analysis, we recommend the use of the full likelihood in these more exotic scenarios or with new datasets.

\section{Discussion}\label{sec:discussion}
We have demonstrated the analysis procedure that faithfully extracts information from the joint combination of BAO, BBN, and CMB datasets, paying particular attention to the role of various assumptions that may be made about BBN.  Our major innovations include
\begin{enumerate}
    \item A CMB-independent determination of $h$, which includes DESI DR2 and a full BBN likelihood, in both $\Lambda$CDM and $\Lambda$CDM+$N_{\rm{eff}}$ (figure~\ref{fig:BAO_BBN}; eqs.~\eqref{eq:h_BAO_BBN},~\eqref{eq:h_Neff_BAO_BBN}),
    \item A BAO+BBN+CMB joint analysis with a full BBN likelihood, in both $\Lambda$CDM and $\Lambda$CDM+$N_{\rm{eff}}$, and joint determination of all cosmological parameters (figures~\ref{fig:all},~\ref{fig:h_all}; eqs.~\eqref{eq:h_all},~\eqref{eq:h_neff_full}),
    \item Demonstration of the BBN reaction network as a systematic in determining both $h$ and $N_{\rm{eff}}$ in these data combinations (figures~\ref{fig:BAO_BBN},~\ref{fig:h_all},~\ref{fig:full_old_new},~\ref{fig:BBN_CMB}; eqs.~\eqref{eq:h_all},~\eqref{eq:h_neff_full}),
    \item A comparison of methods to summarize the BBN likelihood against analyses using the full likelihood (figure~\ref{fig:prior_vs_full}).
\end{enumerate}

Following on the results and discussions in this manuscript, we have the following suggestions and recommendations for future analyses involving BBN:
\begin{enumerate}
    \item We encourage the use of the deuterium measurement from ref.~\cite{Cooke_2018}, as this is a self-consistent reanalysis of available data rather than a simple average.  We encourage the use of the helium-4 measurement from ref.~\cite{aver_2026}, as, among other reasons, this reference eliminates a major systematic (linear regression) present in other determinations of the primordial helium-4 abundance.
    \item We recommend the use of the PRIMAT 2023 reaction network or a network that incorporates the deuterium reaction rates derived via Gaussian process regression in ref.~\cite{launders2026}\footnote{These rates are tabulated in the \href{https://github.com/cgiovanetti/LINX/tree/main/linx/data/nuclear_rates/key_recommended}{LINX \texttt{key\_recommended} network}.} (which shows good agreement in the predicted D/H with the PRIMAT 2023 network in both central value and uncertainty).  Should future analyses use a reaction network other than these, the effects on both $N_{\rm{eff}}$ and $\omega_b$ should be understood and acknowledged.  Ref.~\cite{launders2026} found that the method used to obtain the PArthENoPE rates leads to a bias in the recovered value of D/H, and therefore we do not recommend the use of the PArthENoPE reaction network.  We note the uncertainties in D/H obtained by PRyM/YOF are significantly larger than those obtained by both PRIMAT 2023 and the Gaussian process method in ref.~\cite{launders2026}.
    \item We suggest the use of a full BBN likelihood in joint analyses, similar to that in eq.~\eqref{eq:llBBN}, for authoritative analyses or in applications where precision is required.  We do not recommend the use of a Gaussian prior on $\omega_b$, especially in analyses where $N_{\rm{eff}}$ or other new physics parameters that may correlate with BBN abundances are allowed to float.  We caution that analyses that summarize BBN posteriors with Gaussian priors on D/H should be validated in scenarios outside the limited analyses performed here.
\end{enumerate}
This third recommendation is the least often followed in the existing literature.  As we have shown, these approximations are largely benign in $\Lambda$CDM and $\Lambda$CDM+$N_{\rm eff}$.  However, we have demonstrated above that our recommended procedure is well within reach technologically.  Summaries and priors beyond the two models considered here need to be validated and require running a full analysis, and in new physics analyses where the number of new physics parameters grows the resource requirements of a summarized analysis catch up quickly to those of a full analysis (indeed in $\Lambda$CDM they are already close).  Meanwhile these summary procedures risk shifts in estimated means and uncertainties, especially in cosmologies involving new physics and applications that demand high precision.  They are more difficult to keep up to date, as priors must be recomputed \textit{from full analyses} as new nuclear physics measurements and astrophysical observations become available.  As discussed in section~\ref{sec:intro}, the proliferation of minor tensions across cosmological epochs makes these small shifts from failing to use a full likelihood potentially important.  The use of a full likelihood is, in our view, the approach that is most consistent with the enterprise of precision cosmology. 

\begin{figure}
    \centering
    \includegraphics[width=0.7\linewidth]{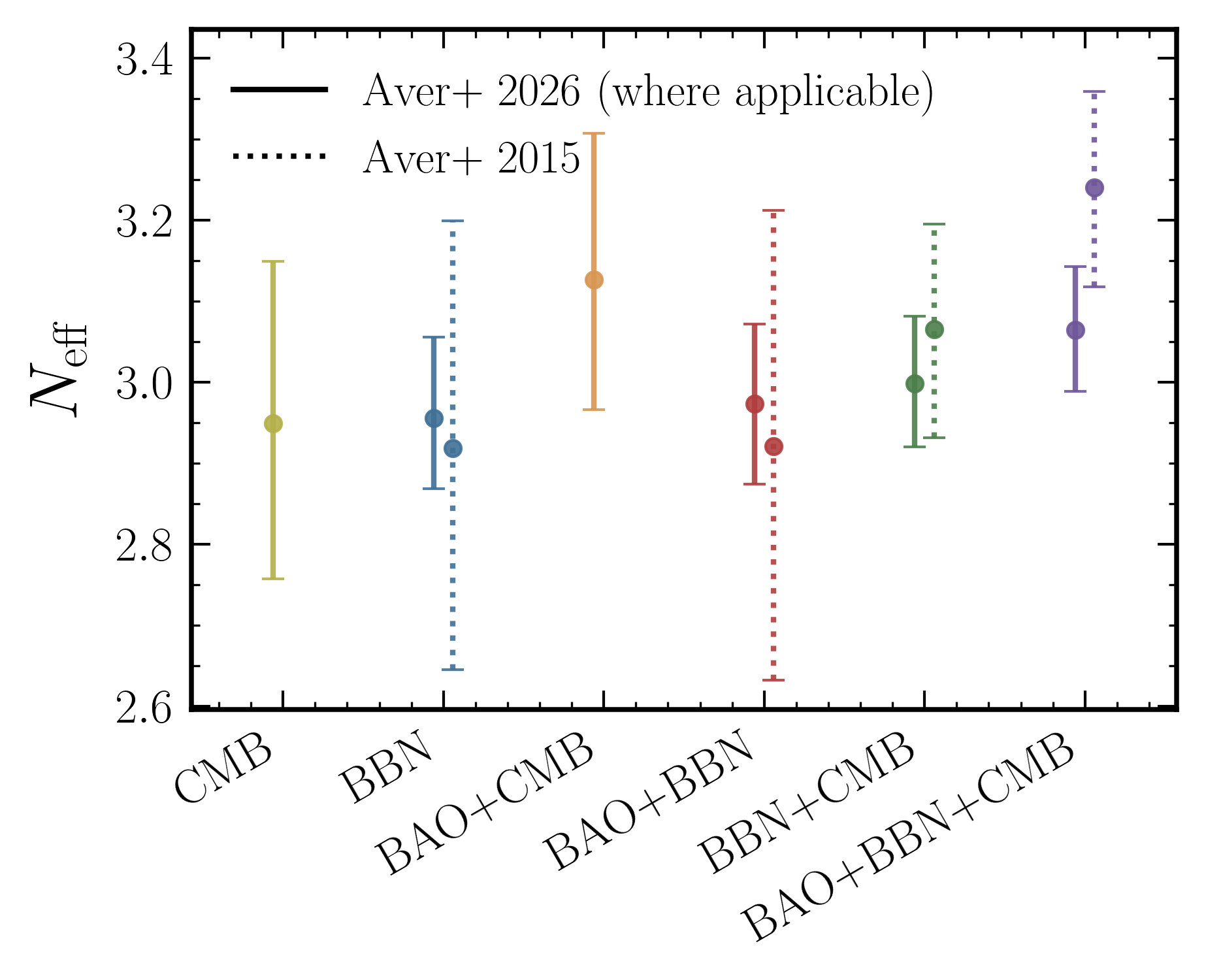}
    \caption{Fiducial $N_{\rm{eff}}$ from all major analyses performed in this work in $\Lambda$CDM+$N_{\rm{eff}}$.  Where applicable, we also show the equivalent results using the older Y$_{\rm{P}}$ measurement from ref.~\cite{Aver_2015} as dotted lines to illustrate the impact of the more recent measurements in ref.~\cite{aver_2026}.}
    \label{fig:Neff}
\end{figure}

We summarize our results for $N_{\rm{eff}}$ from our recommended analysis procedure in Figure~\ref{fig:Neff}, including the older Y$_{\rm{P}}$ measurement from ref.~\cite{Aver_2015} for comparison where applicable.  We find that the combination of BBN, BAO and CMB data tightly constrains $N_{\rm{eff}}$, and we expect this sensitivity to carry over into analyses of other physics beyond $\Lambda$CDM and beyond the Standard Model.

We also note the unusually high $N_{\rm{eff}}$ preferred by BAO+BBN+CMB when the older, less precise measurement of primordial helium-4 is used, where permissive BBN constraints on $N_{\rm{eff}}$ combine with the joint PRIMAT+CMB preference for high $N_{\rm{eff}}$ to yield $N_{\rm{eff}}=3.24\pm0.12$.  We highlight this result as a final cautionary example: a less judicious selection of BBN data might have produced an(other) apparent $\sim 2\sigma$ tension with $\Lambda$CDM.  

\section*{Data Availability}

The nested sampling results for the analyses discussed in this manuscript are available on Zenodo at https://doi.org/10.5281/zenodo.21378961.

\section*{Acknowledgments}
We thank Sam Goldstein for helpful discussions, and Tim Launders for assistance in correcting an error in the PRyM/YOF reaction network.
M.A. was supported by the Boston University Undergraduate Research Opportunities Program. 
N.D. was supported by the Boston University Society of Fellows. 
C.G. is supported by the Office of High Energy Physics of the U.S. Department of Energy under contract DE-AC02-05CH11231. 
H.L. is supported by the U.S. Department of Energy under grants DE-SC0026297 and DE-SC0013895. 
In addition, H.L. is supported by the Cecile K. Dalton Career Development Professorship, endowed by Boston University trustee Nathaniel Dalton and Amy Gottleib Dalton.

\appendix
\section{Complete Results}\label{app:results}
Here we present in full the posteriors obtained for cosmological parameters for all of our analyses. The results contain various combinations of BBN, BAO, and CMB data.  
These analyses are displayed for each of the three BBN networks discussed in this manuscript. We only present results that incorporate BBN data as this is the novel element introduced in our project. 

\begin{table*}[ht]
\centering
\footnotesize
\renewcommand{\arraystretch}{1.25}
\begin{tabular}{|l|c|c|c|c|c|}
\hline
 & \shortstack{BBN} & \shortstack{BBN+BAO} & \shortstack{BBN+CMB} & \shortstack{BBN+BAO+CMB} \\
\hline
100$\omega_{b}$ & $2.193_{-0.026}^{+0.022}$ & $2.190_{-0.023}^{+0.023}$ & $2.219_{-0.012}^{+0.012}$ & $2.236_{-0.011}^{+0.011}$\\
 \hline 
$\Omega_{m}$ &  & $0.2991_{-0.011}^{+0.0095}$ & $0.3217_{-0.0077}^{+0.0070}$ & $0.3023_{-0.0037}^{+0.0034}$\\
 \hline 
$h$ &  & $0.6833_{-0.0053}^{+0.0048}$ & $0.6683_{-0.0047}^{+0.0054}$ & $0.6823_{-0.0026}^{+0.0027}$\\
 \hline 
$\ln(10^{10}A_s)$ &  &  & $3.041_{-0.014}^{+0.015}$ & $3.042_{-0.014}^{+0.016}$\\
 \hline 
$n_{s}$ &  &  & $0.9606_{-0.0036}^{+0.0040}$ & $0.9689_{-0.0033}^{+0.0033}$\\
 \hline 
$100\tau_{\mathrm{reio}}$ &  &  & $5.21_{-0.69}^{+0.71}$ & $5.66_{-0.72}^{+0.77}$\\
 \hline 
\hline
\end{tabular}

\caption{Central values and uncertainties of cosmological parameters from posteriors of variations of BAO, BBN, and CMB, $\Lambda \mathrm{CDM}$ analyses. The BBN analysis uses the PRIMAT network.}
\label{tab:P23_full}
\end{table*}

\begin{table*}[ht]
\centering
\footnotesize
\renewcommand{\arraystretch}{1.25}

\begin{tabular}{|l|c|c|c|c|c|}
\hline
 & \shortstack{BBN} & \shortstack{BBN+BAO} & \shortstack{BBN+CMB} & \shortstack{BBN+BAO+CMB} \\
\hline
100$\omega_{b}$ & $2.230_{-0.028}^{+0.024}$ & $2.227_{-0.025}^{+0.026}$ & $2.230_{-0.013}^{+0.013}$ & $2.246_{-0.0094}^{+0.011}$\\
 \hline 
$\Omega_{m}$ &  & $0.2983_{-0.010}^{+0.0098}$ & $0.3173_{-0.0080}^{+0.0082}$ & $0.3008_{-0.0033}^{+0.0034}$\\
 \hline 
$h$ &  & $0.6865_{-0.0048}^{+0.0052}$ & $0.6719_{-0.0057}^{+0.0056}$ & $0.6838_{-0.0025}^{+0.0027}$\\
 \hline 
$\ln(10^{10}A_s)$ &  &  & $3.042_{-0.016}^{+0.016}$ & $3.045_{-0.015}^{+0.015}$\\
 \hline 
$n_{s}$ &  &  & $0.9626_{-0.0040}^{+0.0043}$ & $0.9692_{-0.0033}^{+0.0030}$\\
 \hline 
$100\tau_{\mathrm{reio}}$ &  &  & $5.35_{-0.70}^{+0.76}$ & $5.73_{-0.75}^{+0.77}$\\
 \hline 
\hline
\end{tabular}

\caption{Central values and uncertainties of cosmological parameters from posteriors of variations of BAO, BBN, and CMB, $\Lambda \mathrm{CDM}$ analyses. The BBN analysis uses the PArthENoPE network.}
\label{tab:PA_full}
\end{table*}

\begin{table*}[ht]
\centering
\footnotesize
\renewcommand{\arraystretch}{1.25}

\begin{tabular}{|l|c|c|c|c|c|}
\hline
 & \shortstack{BBN} & \shortstack{BBN+BAO} & \shortstack{BBN+CMB} & \shortstack{BBN+BAO+CMB} \\
\hline
100$\omega_{b}$ & $2.231_{-0.055}^{+0.057}$ & $2.220_{-0.052}^{+0.050}$ & $2.232_{-0.014}^{+0.015}$ & $2.247_{-0.013}^{+0.012}$\\
 \hline 
$\Omega_{m}$ &  & $0.297_{-0.010}^{+0.011}$ & $0.3168_{-0.0080}^{+0.0081}$ & $0.3011_{-0.0035}^{+0.0037}$\\
 \hline 
$h$ &  & $0.6858_{-0.0068}^{+0.0067}$ & $0.6723_{-0.0058}^{+0.0058}$ & $0.6836_{-0.0029}^{+0.0028}$\\
 \hline  
$\ln(10^{10}A_s)$ &  &  & $3.043_{-0.015}^{+0.015}$ & $3.044_{-0.016}^{+0.016}$\\
 \hline 
$n_{s}$ &  &  & $0.9625_{-0.0045}^{+0.0044}$ & $0.9691_{-0.0033}^{+0.0032}$\\
 \hline 
$100\tau_{\mathrm{reio}}$ &  &  & $5.41_{-0.77}^{+0.71}$ & $5.73_{-0.75}^{+0.79}$\\
 \hline 
\hline
\end{tabular}

\caption{Central values and uncertainties of cosmological parameters from posteriors of variations of BAO, BBN, and CMB, $\Lambda \mathrm{CDM}$ analyses. The BBN analysis uses the PRyM/YOF network.}
\label{tab:YOF_full}
\end{table*}

\begin{table*}[ht]
\centering
\footnotesize
\renewcommand{\arraystretch}{1.25}

\begin{tabular}{|l|c|c|c|}
\hline
 & \shortstack{CMB} & \shortstack{BAO+CMB} \\
\hline
100$\omega_{b}$ & $2.224_{-0.023}^{+0.022}$ & $2.256_{-0.016}^{+0.016}$\\
 \hline 
$\Omega_{m}$ & $0.319_{-0.010}^{+0.011}$ & $0.2994_{-0.0040}^{+0.0041}$\\
 \hline 
$h$ & $0.666_{-0.015}^{+0.015}$ & $0.689_{-0.0097}^{+0.010}$\\
 \hline 
$N_{\rm{eff}}$ & $2.95_{-0.19}^{+0.20}$ & $3.13_{-0.16}^{+0.18}$\\
 \hline 
$\ln(10^{10}A_s)$ & $3.038_{-0.018}^{+0.019}$ & $3.047_{-0.018}^{+0.018}$\\
 \hline 
$n_{s}$ & $0.9593_{-0.0087}^{+0.0087}$ & $0.9717_{-0.0059}^{+0.0059}$\\
 \hline 
$100\tau_{\mathrm{reio}}$ & $5.35_{-0.76}^{+0.75}$ & $5.76_{-0.78}^{+0.79}$\\
 \hline 
\hline
\end{tabular}

\caption{Central values and uncertainties of cosmological parameters from posteriors of variations of BAO and CMB, $\Lambda \mathrm{CDM} + N_{\text{eff}}$ analyses. Relies on the DESI DR2 and Planck 2018 likelihoods.}
\label{tab:CMB_BAO_Neff}
\end{table*}

\begin{table*}[ht]
\centering
\footnotesize
\renewcommand{\arraystretch}{1.25}

\begin{tabular}{|l|c|c|c|c|c|}
\hline
 & \shortstack{BBN} & \shortstack{BAO+BBN} & \shortstack{BBN+CMB} & \shortstack{BAO+BBN+CMB} \\
\hline
100$\omega_{b}$ & $2.177_{-0.024}^{+0.026}$ & $2.179_{-0.026}^{+0.025}$ & $2.214_{-0.014}^{+0.015}$ & $2.238_{-0.013}^{+0.013}$\\
 \hline 
$\Omega_{m}$ &  & $0.2973_{-0.0083}^{+0.0093}$ & $0.3235_{-0.0084}^{+0.0083}$ & $0.3017_{-0.0037}^{+0.0038}$\\
 \hline 
$h$ &  & $0.6791_{-0.0075}^{+0.0069}$ & $0.6651_{-0.0078}^{+0.0079}$ & $0.6838_{-0.0054}^{+0.0052}$\\
 \hline 
$N_{\rm{eff}}$ & $2.96_{-0.087}^{+0.10}$ & $2.974_{-0.099}^{+0.098}$ & $2.999_{-0.079}^{+0.083}$ & $3.065_{-0.076}^{+0.078}$\\
 \hline 
$\ln(10^{10}A_s)$ &  &  & $3.038_{-0.015}^{+0.016}$ & $3.043_{-0.015}^{+0.017}$\\
 \hline 
$n_{s}$ &  &  & $0.9595_{-0.0049}^{+0.0051}$ & $0.9695_{-0.0040}^{+0.0040}$\\
 \hline 
$100\tau_{\mathrm{reio}}$ &  &  & $5.21_{-0.72}^{+0.74}$ & $5.67_{-0.72}^{+0.74}$\\
 \hline 
\hline
\end{tabular}

\caption{Central values and uncertainties of cosmological parameters from posteriors of variations of BAO, BBN, and CMB, $\Lambda \mathrm{CDM} + N_{\text{eff}}$ analyses. The BBN analysis uses the PRIMAT network.}
\label{tab:P23neff_full}
\end{table*}

\begin{table*}[ht]
\centering
\footnotesize
\renewcommand{\arraystretch}{1.25}

\begin{tabular}{|l|c|c|c|c|c|}
\hline
 & \shortstack{BBN} & \shortstack{BAO+BBN} & \shortstack{BBN+CMB} & \shortstack{BBN+BAO+CMB} \\
\hline
100$\omega_{b}$ & $2.215_{-0.034}^{+0.032}$ & $2.217_{-0.033}^{+0.030}$ & $2.221_{-0.014}^{+0.015}$ & $2.244_{-0.012}^{+0.012}$\\
 \hline 
$\Omega_{m}$ &  & $0.297_{-0.0099}^{+0.010}$ & $0.3203_{-0.0087}^{+0.0087}$ & $0.3013_{-0.0037}^{+0.0038}$\\
 \hline 
$h$ &  & $0.6814_{-0.0090}^{+0.0083}$ & $0.6659_{-0.0080}^{+0.0078}$ & $0.6825_{-0.0052}^{+0.0051}$\\
 \hline 
$N_{\rm{eff}}$ & $2.94_{-0.10}^{+0.13}$ & $2.97_{-0.11}^{+0.11}$ & $2.960_{-0.079}^{+0.079}$ & $3.027_{-0.079}^{+0.079}$\\
 \hline 
$\ln(10^{10}A_s)$ &  &  & $3.039_{-0.016}^{+0.015}$ & $3.043_{-0.015}^{+0.017}$\\
 \hline 
$n_{s}$ &  &  & $0.9592_{-0.0053}^{+0.0051}$ & $0.9684_{-0.0039}^{+0.0040}$\\
 \hline 
$100\tau_{\mathrm{reio}}$ &  &  & $5.30_{-0.74}^{+0.73}$ & $5.74_{-0.71}^{+0.76}$\\
 \hline 
\hline
\end{tabular}

\caption{Central values and uncertainties of cosmological parameters from posteriors of variations of BAO, BBN, and CMB, $\Lambda \mathrm{CDM} + N_{\text{eff}}$ analyses. The BBN analysis uses the PArthENoPE network.}
\label{tab:PAneff_full}
\end{table*}

\begin{table*}[ht]
\centering
\footnotesize
\renewcommand{\arraystretch}{1.25}

\begin{tabular}{|l|c|c|c|c|c|}
\hline
 & \shortstack{BBN} & \shortstack{BAO+BBN} & \shortstack{BBN+CMB} & \shortstack{BBN+BAO+CMB} \\
\hline
100$\omega_{b}$ & $2.211_{-0.055}^{+0.055}$ & $2.212_{-0.058}^{+0.057}$ & $2.223_{-0.015}^{+0.015}$ & $2.246_{-0.013}^{+0.013}$\\
 \hline 
$\Omega_{m}$ &  & $0.2983_{-0.0083}^{+0.0091}$ & $0.3195_{-0.0079}^{+0.0082}$ & $0.3010_{-0.0035}^{+0.0037}$\\
 \hline 
$h$ &  & $0.6811_{-0.0080}^{+0.0081}$ & $0.6659_{-0.0075}^{+0.0076}$ & $0.6812_{-0.0052}^{+0.0054}$\\
 \hline 
$N_{\rm{eff}}$ & $2.96_{-0.098}^{+0.10}$ & $2.967_{-0.095}^{+0.094}$ & $2.950_{-0.078}^{+0.080}$ & $2.991_{-0.080}^{+0.081}$\\
 \hline 
$\ln(10^{10}A_s)$ &  &  & $3.039_{-0.014}^{+0.015}$ & $3.041_{-0.015}^{+0.015}$\\
 \hline 
$n_{s}$ &  &  & $0.9591_{-0.0052}^{+0.0053}$ & $0.9675_{-0.0039}^{+0.0041}$\\
 \hline 
$100\tau_{\mathrm{reio}}$ &  &  & $5.34_{-0.68}^{+0.69}$ & $5.69_{-0.73}^{+0.74}$\\
 \hline 
\hline
\end{tabular}

\caption{Central values and uncertainties of cosmological parameters from posteriors of variations of BAO, BBN, and CMB, $\Lambda \mathrm{CDM} + N_{\text{eff}}$ analyses. The BBN analysis uses the PRyM/YOF network.}
\label{tab:YOFneff_full}
\end{table*}

\bibliographystyle{jcap}
\bibliography{references}

@article{LINXlong,
   title={Fast, differentiable, and extensible big bang nucleosynthesis package},
   volume={112},
   ISSN={2470-0029},
   url={http://dx.doi.org/10.1103/f3tj-r882},
   DOI={10.1103/f3tj-r882},
   number={6},
   journal={Physical Review D},
   publisher={American Physical Society (APS)},
   author={Giovanetti, Cara and Lisanti, Mariangela and Liu, Hongwan and Mishra-Sharma, Siddharth and Ruderman, Joshua T.},
   year={2025},
   month=sep }

@article{Poulin:2026ltf,
    author = "Poulin, Vivian and Froustey, Julien and Pitrou, Cyril and Smith, Tristan L.",
    title = "{What could an emerging Big Bang Nucleosynthesis discrepancy be hinting at?}",
    eprint = "2607.20635",
    archivePrefix = "arXiv",
    primaryClass = "astro-ph.CO",
    month = "7",
    year = "2026"
}

@article{Elbers:2025vlz,
    author = "Elbers, W. and others",
    title = "{Constraints on neutrino physics from DESI DR2 BAO and DR1 full shape}",
    eprint = "2503.14744",
    archivePrefix = "arXiv",
    primaryClass = "astro-ph.CO",
    reportNumber = "FERMILAB-PUB-25-0168-PPD",
    doi = "10.1103/w9pk-xsk7",
    journal = "Phys. Rev. D",
    volume = "112",
    number = "8",
    pages = "083513",
    year = "2025"
}

@article{Tait:2026vrd,
    author = "Tait, Tim M. P.",
    title = "{Deuterium Production in an Effective Field Theory Constructed from On-Shell Amplitudes}",
    eprint = "2608.23699",
    archivePrefix = "arXiv",
    primaryClass = "nucl-th",
    reportNumber = "UCI-HEP-TR-2026-10",
    month = "8",
    year = "2026"
}

@article{Tait:2026guk,
    author = "Tait, Tim M. P.",
    title = "{Deuterium-Proton Fusion in an Effective Field Theory Constructed from On-Shell Amplitudes}",
    eprint = "2607.05514",
    archivePrefix = "arXiv",
    primaryClass = "nucl-th",
    month = "7",
    year = "2026"
}

@article{Cooke_2018,
   title={One Percent Determination of the Primordial Deuterium Abundance},
   volume={855},
   ISSN={1538-4357},
   url={http://dx.doi.org/10.3847/1538-4357/aaab53},
   DOI={10.3847/1538-4357/aaab53},
   number={2},
   journal={The Astrophysical Journal},
   publisher={American Astronomical Society},
   author={Cooke, Ryan J. and Pettini, Max and Steidel, Charles C.},
   year={2018},
   month=mar, pages={102} }

@misc{aver_2026,
      title={The {LBT} {Y}$_\mathrm{p}$ Project {IV}: A New Value of the Primordial Helium Abundance}, 
      author={Erik Aver and Evan D. Skillman and Richard W. Pogge and Noah S. J. Rogers and Miqaela K. Weller and Keith A. Olive and Danielle A. Berg and John J. Salzer and John H. Miller Jr. and José Eduardo Méndez-Delgado},
      year={2026},
      eprint={2601.22238},
      archivePrefix={arXiv},
      primaryClass={astro-ph.CO},
      url={https://arxiv.org/abs/2601.22238}, 
}

@article{LINX_short,
   title={Cosmological parameter estimation with a joint-likelihood analysis of the cosmic microwave background and big bang nucleosynthesis},
   volume={112},
   ISSN={2470-0029},
   url={http://dx.doi.org/10.1103/wspy-s948},
   DOI={10.1103/wspy-s948},
   number={6},
   journal={Physical Review D},
   publisher={American Physical Society (APS)},
   author={Giovanetti, Cara and Lisanti, Mariangela and Liu, Hongwan and Mishra-Sharma, Siddharth and Ruderman, Joshua T.},
   year={2025},
   month=sep }

@misc{launders2026,
      title={A data-driven prediction for the primordial deuterium abundance}, 
      author={Timothy Launders and Cara Giovanetti and Hongwan Liu},
      year={2026},
      eprint={2604.16600},
      archivePrefix={arXiv},
      primaryClass={astro-ph.CO},
      url={https://arxiv.org/abs/2604.16600}, 
}

@INPROCEEDINGS{Skilling_2004,
       author = {{Skilling}, John},
        title = "{Nested Sampling}",
    booktitle = {{Bayesian} Inference and Maximum Entropy Methods in Science and Engineering: 24th International Workshop on {Bayesian} Inference and Maximum Entropy Methods in Science and Engineering},
         year = 2004,
       editor = {{Fischer}, Rainer and {Preuss}, Roland and {Toussaint}, Udo Von},
       series = {American Institute of Physics Conference Series},
       volume = {735},
        month = nov,
    publisher = {AIP},
        pages = {395-405},
          doi = {10.1063/1.1835238},
       adsurl = {https://ui.adsabs.harvard.edu/abs/2004AIPC..735..395S}
}

@article{Skilling_2006,
author = {John Skilling},
title = {{Nested sampling for general {Bayesian} computation}},
volume = {1},
journal = {{Bayesian} Analysis},
number = {4},
publisher = {International Society for {Bayesian} Analysis},
pages = {833 -- 859},
year = {2006},
doi = {10.1214/06-BA127},
URL = {https://doi.org/10.1214/06-BA127}
}

@ARTICLE{Speagle_2020,
       author = {{Speagle}, Joshua S.},
        title = "{DYNESTY: a dynamic nested sampling package for estimating {Bayesian} posteriors and evidences}",
      journal = {\mnras},
         year = 2020,
        month = apr,
       volume = {493},
       number = {3},
        pages = {3132-3158},
          doi = {10.1093/mnras/staa278},
archivePrefix = {arXiv},
       eprint = {1904.02180},
 primaryClass = {astro-ph.IM},
       adsurl = {https://ui.adsabs.harvard.edu/abs/2020MNRAS.493.3132S}
}

@software{speagle_2023,
  title = {dynesty: v2.1.3},
  author = {Koposov, Sergey and Speagle, Josh and Barbary, Kyle and Ashton, Gregory and Bennett, Ed and Buchner, Johannes and Scheffler, Carl and Cook, Ben and Talbot, Colm and Guillochon, James and Cubillos, Patricio and Asensio Ramos, Andrés and Johnson, Ben and Lang, Dustin and Ilya and Dartiailh, Matthieu and Nitz, Alex and McCluskey, Andrew and Archibald, Anne},
  month = oct,
  year = 2023,
  version = {v2.1.3},
  doi = {10.5281/zenodo.8408702},
  url = {https://doi.org/10.5281/zenodo.8408702},
  publisher = {Zenodo},
  address = {Geneva},
  type = {software},
  license = {other-open},
  note = {Version 2.1.3. License type: other-open.}
}

@misc{yeh2026,
      title={{The LBT $Y_{\rm p}$ Project V: Cosmological Implications of a New Determination of Primordial $^4$He}}, 
      author={Tsung-Han Yeh and Keith A. Olive and Brian D. Fields and Erik Aver and Richard W. Pogge and Noah S. J. Rogers and Evan D. Skillman and Miqaela K. Weller},
      year={2026},
      eprint={2601.22239},
      archivePrefix={arXiv},
      primaryClass={astro-ph.CO},
      url={https://arxiv.org/abs/2601.22239}, 
}

@article{Planck2018,
   title={{Planck2018 results: VI. Cosmological parameters}},
   volume={641},
   ISSN={1432-0746},
   url={http://dx.doi.org/10.1051/0004-6361/201833910},
   DOI={10.1051/0004-6361/201833910},
   journal={Astronomy \& Astrophysics},
   publisher={EDP Sciences},
   author={Aghanim, N. and others},
   year={2020},
   month=sep, pages={A6} }

@article{Pitrou_2018,
   author = {Pitrou, Cyril and Coc, Alain and Uzan, Jean-Philippe and Vangioni, Elisabeth},
   title = {Precision big bang nucleosynthesis with improved Helium-4 predictions},
   journal = {Submitted to Phys. Rept.},
   archivePrefix = "arXiv",
   eprint = {1801.08023},
   year = 2018
   }

@article{Consiglio_2018,
   title={{PArthENoPE} reloaded},
   volume={233},
   ISSN={0010-4655},
   url={http://dx.doi.org/10.1016/j.cpc.2018.06.022},
   DOI={10.1016/j.cpc.2018.06.022},
   journal={Computer Physics Communications},
   publisher={Elsevier BV},
   author={Consiglio, R. and de Salas, P.F. and Mangano, G. and Miele, G. and Pastor, S. and Pisanti, O.},
   year={2018},
   month=dec, pages={237–242} }

@article{Gariazzo:2021iiu,
    author = "Gariazzo, S. and F. de Salas, P. and Pisanti, O. and Consiglio, R.",
    title = "{PArthENoPE revolutions}",
    eprint = "2103.05027",
    archivePrefix = "arXiv",
    primaryClass = "astro-ph.IM",
    doi = "10.1016/j.cpc.2021.108205",
    journal = "Comput. Phys. Commun.",
    volume = "271",
    pages = "108205",
    year = "2022"
}

@article{Pisanti_2021,
   title={Primordial Deuterium after {LUNA}: concordances and error budget},
   volume={2021},
   ISSN={1475-7516},
   url={http://dx.doi.org/10.1088/1475-7516/2021/04/020},
   DOI={10.1088/1475-7516/2021/04/020},
   number={04},
   journal={Journal of Cosmology and Astroparticle Physics},
   publisher={IOP Publishing},
   author={Pisanti, O. and Mangano, G. and Miele, G. and Mazzella, P.},
   year={2021},
   month=apr, pages={020} }

@misc{calabrese2025atacamacosmologytelescopedr6,
      title={{The Atacama Cosmology Telescope: DR6 Constraints on Extended Cosmological Models}}, 
      author={Erminia Calabrese and others},
      collaboration={Atacama Cosmology Telescope},
      year={2025},
      eprint={2503.14454},
      archivePrefix={arXiv},
      primaryClass={astro-ph.CO},
      url={https://arxiv.org/abs/2503.14454}, 
}

@misc{ferreira2025baocmbtensionimplicationsinflation,
      title={{The BAO-CMB Tension and Implications for Inflation}}, 
      author={Elisa G. M. Ferreira and Evan McDonough and Lennart Balkenhol and Renata Kallosh and Lloyd Knox and Andrei Linde},
      year={2025},
      eprint={2507.12459},
      archivePrefix={arXiv},
      primaryClass={astro-ph.CO},
      url={https://arxiv.org/abs/2507.12459}, 
}

@article{Pitrou_2021,
   title={A new tension in the cosmological model from primordial deuterium?},
   volume={502},
   ISSN={1365-2966},
   url={http://dx.doi.org/10.1093/mnras/stab135},
   DOI={10.1093/mnras/stab135},
   number={2},
   journal={Monthly Notices of the Royal Astronomical Society},
   publisher={Oxford University Press (OUP)},
   author={Pitrou, Cyril and Coc, Alain and Uzan, Jean-Philippe and Vangioni, Elisabeth},
   year={2021},
   month=Jan, pages={2474–2481} }

@article{Adame_2025,
   title={{DESI 2024 VI:  cosmological constraints from the measurements of baryon acoustic oscillations}},
   volume={2025},
   ISSN={1475-7516},
   url={http://dx.doi.org/10.1088/1475-7516/2025/02/021},
   DOI={10.1088/1475-7516/2025/02/021},
   number={02},
   journal={Journal of Cosmology and Astroparticle Physics},
   publisher={IOP Publishing},
   author={Adame, A.G. and others},
   collaboration={DESI},
   year={2025},
   month=Feb, pages={021} }

@misc{craig2024nusgoodnews,
      title={No $\nu$s is Good News}, 
      author={Nathaniel Craig and Daniel Green and Joel Meyers and Surjeet Rajendran},
      year={2024},
      eprint={2405.00836},
      archivePrefix={arXiv},
      primaryClass={astro-ph.CO},
      url={https://arxiv.org/abs/2405.00836}, 
}

@article{Yeh_2021,
   title={The impact of new d(p,$\gamma$)$^3${He} rates on Big Bang Nucleosynthesis},
   volume={2021},
   ISSN={1475-7516},
   url={http://dx.doi.org/10.1088/1475-7516/2021/03/046},
   DOI={10.1088/1475-7516/2021/03/046},
   number={03},
   journal={Journal of Cosmology and Astroparticle Physics},
   publisher={IOP Publishing},
   author={Yeh, Tsung-Han and Olive, Keith A. and Fields, Brian D.},
   year={2021},
   month=Mar, pages={046} }

@article{Xu_2013,
   title={{NACRE II}: an update of the {NACRE} compilation of charged-particle-induced thermonuclear reaction rates for nuclei with mass number $A<16$},
   volume={918},
   ISSN={0375-9474},
   url={http://dx.doi.org/10.1016/j.nuclphysa.2013.09.007},
   DOI={10.1016/j.nuclphysa.2013.09.007},
   journal={Nuclear Physics A},
   publisher={Elsevier BV},
   author={Xu, Y. and Takahashi, K. and Goriely, S. and Arnould, M. and Ohta, M. and Utsunomiya, H.},
   year={2013},
   month=Nov, pages={61–169} }

@misc{burns2023prymordialminutesstandardmodel,
      title={{PRyMordial}: The First Three Minutes, Within and Beyond the Standard Model}, 
      author={Anne-Katherine Burns and Tim M. P. Tait and Mauro Valli},
      year={2023},
      eprint={2307.07061},
      archivePrefix={arXiv},
      primaryClass={hep-ph},
      url={https://arxiv.org/abs/2307.07061}, 
}

@misc{lesgourgues_2011a,
      title={{The Cosmic Linear Anisotropy Solving System} ({CLASS}) {I}: Overview}, 
      author={Julien Lesgourgues},
      year={2011},
      eprint={1104.2932},
      archivePrefix={arXiv},
      primaryClass={astro-ph.IM}
}

@misc{lesgourgues_2011c,
    author = "Lesgourgues, Julien",
    title = "{The Cosmic Linear Anisotropy Solving System ({CLASS}) {III}: Comparision with {CAMB} for Lambda{CDM}}",
    eprint = "1104.2934",
    archivePrefix = "arXiv",
    primaryClass = "astro-ph.CO",
    reportNumber = "CERN-PH-TH-2011-083, LAPTH-011-11",
    month = "4",
    year = "2011"
}

@article{lesgourgues_2011d,
   title={{The Cosmic Linear Anisotropy Solving System} ({CLASS}) {IV}: efficient implementation of non-cold relics},
   volume={2011},
   ISSN={1475-7516},
   url={http://dx.doi.org/10.1088/1475-7516/2011/09/032},
   DOI={10.1088/1475-7516/2011/09/032},
   number={09},
   journal={Journal of Cosmology and Astroparticle Physics},
   publisher={IOP Publishing},
   author={Lesgourgues, Julien and Tram, Thomas},
   year={2011},
   month=sep, pages={032–032} }

@article{Planck_2018_likelihoods,
   title={Planck2018 results: {V}. {CMB} power spectra and likelihoods},
   volume={641},
   ISSN={1432-0746},
   url={http://dx.doi.org/10.1051/0004-6361/201936386},
   DOI={10.1051/0004-6361/201936386},
   journal={Astronomy \& Astrophysics},
   publisher={EDP Sciences},
   author={Aghanim, N. and others},
   year={2020},
   month=sep, pages={A5} }

@article{Planck2013,
   title={Planck2013 results. {XVI}. Cosmological parameters},
   volume={571},
   ISSN={1432-0746},
   url={http://dx.doi.org/10.1051/0004-6361/201321591},
   DOI={10.1051/0004-6361/201321591},
   journal={Astronomy \& Astrophysics},
   publisher={EDP Sciences},
   author={Ade, P. A. R. and others},
   year={2014},
   month=oct, pages={A16} }

@article{Fixsen_2009,
   title={THE TEMPERATURE OF THE COSMIC MICROWAVE BACKGROUND},
   volume={707},
   ISSN={1538-4357},
   url={http://dx.doi.org/10.1088/0004-637X/707/2/916},
   DOI={10.1088/0004-637x/707/2/916},
   number={2},
   journal={The Astrophysical Journal},
   publisher={American Astronomical Society},
   author={Fixsen, D. J.},
   year={2009},
   month=Nov, pages={916–920} }

@article{Aver_2015,
   title={The effects of {He}{I} \ensuremath{\lambda}10830 on helium abundance determinations},
   volume={2015},
   ISSN={1475-7516},
   url={http://dx.doi.org/10.1088/1475-7516/2015/07/011},
   DOI={10.1088/1475-7516/2015/07/011},
   number={07},
   journal={Journal of Cosmology and Astroparticle Physics},
   publisher={IOP Publishing},
   author={Aver, Erik and Olive, Keith A. and Skillman, Evan D.},
   year={2015},
   month=jul, pages={011–011} }

@article{Camphuis_2026,
   title={{SPT-3G D1: CMB temperature and polarization power spectra and cosmology from 2019 and 2020 observations of the SPT-3G main field}},
   volume={113},
   collaboration={{SPT}},
   ISSN={2470-0029},
   url={http://dx.doi.org/10.1103/7wt3-9v2y},
   DOI={10.1103/7wt3-9v2y},
   number={8},
   journal={Physical Review D},
   publisher={American Physical Society (APS)},
   author={Camphuis, E. and others},
   year={2026},
   month=Apr }

@misc{goldstein20262determinationnrmeff,
      title={{A 2\% determination of $N_{\rm eff}$ from primordial element abundance, cosmic microwave background, and baryon acoustic oscillation measurements}}, 
      author={Samuel Goldstein and J. Colin Hill},
      year={2026},
      eprint={2603.13226},
      archivePrefix={arXiv},
      primaryClass={astro-ph.CO},
      url={https://arxiv.org/abs/2603.13226}, 
}

@misc{escudero2026doesnrmeff,
      title={What does it take to have {$N_{\rm eff} < 3$} at {CMB} times?}, 
      author={Miguel Escudero and Maksym Ovchynnikov and Neal Weiner},
      year={2026},
      eprint={2603.22391},
      archivePrefix={arXiv},
      primaryClass={hep-ph},
      url={https://arxiv.org/abs/2603.22391}, 
}

@article{Akita:2020szl,
    author = "Akita, Kensuke and Yamaguchi, Masahide",
    title = "{A precision calculation of relic neutrino decoupling}",
    eprint = "2005.07047",
    archivePrefix = "arXiv",
    primaryClass = "hep-ph",
    doi = "10.1088/1475-7516/2020/08/012",
    journal = "JCAP",
    volume = "08",
    pages = "012",
    year = "2020"
}

@article{Froustey:2020mcq,
    author = "Froustey, Julien and Pitrou, Cyril and Volpe, Maria Cristina",
    title = "{Neutrino decoupling including flavour oscillations and primordial nucleosynthesis}",
    eprint = "2008.01074",
    archivePrefix = "arXiv",
    primaryClass = "hep-ph",
    doi = "10.1088/1475-7516/2020/12/015",
    journal = "JCAP",
    volume = "12",
    pages = "015",
    year = "2020"
}

@article{Bennett:2020zkv,
    author = "Bennett, Jack J. and Buldgen, Gilles and De Salas, Pablo F. and Drewes, Marco and Gariazzo, Stefano and Pastor, Sergio and Wong, Yvonne Y. Y.",
    title = "{Towards a precision calculation of $N_{\rm eff}$ in the Standard Model II: Neutrino decoupling in the presence of flavour oscillations and finite-temperature QED}",
    eprint = "2012.02726",
    archivePrefix = "arXiv",
    primaryClass = "hep-ph",
    reportNumber = "CPPC-2020-10",
    doi = "10.1088/1475-7516/2021/04/073",
    journal = "JCAP",
    volume = "04",
    pages = "073",
    year = "2021"
}

@ARTICLE{Feroz_2009,
       author = {{Feroz}, F. and {Hobson}, M.~P. and {Bridges}, M.},
        title = "{MULTINEST: an efficient and robust {Bayesian} inference tool for cosmology and particle physics}",
      journal = {\mnras},
         year = 2009,
        month = oct,
       volume = {398},
       number = {4},
        pages = {1601-1614},
          doi = {10.1111/j.1365-2966.2009.14548.x},
archivePrefix = {arXiv},
       eprint = {0809.3437},
 primaryClass = {astro-ph},
       adsurl = {https://ui.adsabs.harvard.edu/abs/2009MNRAS.398.1601F}
}

@article{Neal_2003,
author = {Radford M. Neal},
title = {{Slice sampling}},
volume = {31},
journal = {The Annals of Statistics},
number = {3},
publisher = {Institute of Mathematical Statistics},
pages = {705 -- 767},
year = {2003},
doi = {10.1214/aos/1056562461},
URL = {https://doi.org/10.1214/aos/1056562461}
}

@ARTICLE{Handley_2015a,
       author = {{Handley}, W.~J. and {Hobson}, M.~P. and {Lasenby}, A.~N.},
        title = "{polychord: nested sampling for cosmology.}",
      journal = {\mnras},
         year = 2015,
        month = jun,
       volume = {450},
        pages = {L61-L65},
          doi = {10.1093/mnrasl/slv047},
archivePrefix = {arXiv},
       eprint = {1502.01856},
 primaryClass = {astro-ph.CO},
       adsurl = {https://ui.adsabs.harvard.edu/abs/2015MNRAS.450L..61H}
}

@ARTICLE{Handley_2015b,
       author = {{Handley}, W.~J. and {Hobson}, M.~P. and {Lasenby}, A.~N.},
        title = "{POLYCHORD: next-generation nested sampling}",
      journal = {\mnras},
         year = 2015,
        month = nov,
       volume = {453},
       number = {4},
        pages = {4384-4398},
          doi = {10.1093/mnras/stv1911},
archivePrefix = {arXiv},
       eprint = {1506.00171},
 primaryClass = {astro-ph.IM},
       adsurl = {https://ui.adsabs.harvard.edu/abs/2015MNRAS.453.4384H}
}

@article{Cuceu_2019,
   title={{Baryon Acoustic Oscillations and the Hubble constant: past, present and future}},
   volume={2019},
   ISSN={1475-7516},
   url={http://dx.doi.org/10.1088/1475-7516/2019/10/044},
   DOI={10.1088/1475-7516/2019/10/044},
   number={10},
   journal={Journal of Cosmology and Astroparticle Physics},
   publisher={IOP Publishing},
   author={Cuceu, Andrei and Farr, James and Lemos, Pablo and Font-Ribera, Andreu},
   year={2019},
   month=Oct, pages={044–044} }

@article{Addison_2018,
   title={{Elucidating $\Lambda$CDM: Impact of Baryon Acoustic Oscillation Measurements on the Hubble Constant Discrepancy}},
   volume={853},
   ISSN={1538-4357},
   url={http://dx.doi.org/10.3847/1538-4357/aaa1ed},
   DOI={10.3847/1538-4357/aaa1ed},
   number={2},
   journal={The Astrophysical Journal},
   publisher={American Astronomical Society},
   author={Addison, G. E. and Watts, D. J. and Bennett, C. L. and Halpern, M. and Hinshaw, G. and Weiland, J. L.},
   year={2018},
   month=jan, pages={119} }

@article{Schoneberg_2019,
   title={{The BAO+BBN take on the Hubble tension}},
   volume={2019},
   ISSN={1475-7516},
   url={http://dx.doi.org/10.1088/1475-7516/2019/10/029},
   DOI={10.1088/1475-7516/2019/10/029},
   number={10},
   journal={Journal of Cosmology and Astroparticle Physics},
   publisher={IOP Publishing},
   author={Schöneberg, Nils and Lesgourgues, Julien and Hooper, Deanna C.},
   year={2019},
   month=oct, pages={029–029} }

@ARTICLE{smith_1993,
       author = {{Smith}, Michael S. and {Kawano}, Lawrence H. and {Malaney}, Robert A.},
        title = "{Experimental, Computational, and Observational Analysis of Primordial Nucleosynthesis}",
      journal = {\apjs},
         year = 1993,
        month = apr,
       volume = {85},
        pages = {219},
          doi = {10.1086/191763},
       adsurl = {https://ui.adsabs.harvard.edu/abs/1993ApJS...85..219S}
}

@article{Nollett_2000,
   title={Estimating reaction rates and uncertainties for primordial nucleosynthesis},
   volume={61},
   ISSN={1089-4918},
   url={http://dx.doi.org/10.1103/PhysRevD.61.123505},
   DOI={10.1103/physrevd.61.123505},
   number={12},
   journal={Physical Review D},
   publisher={American Physical Society (APS)},
   author={Nollett, Kenneth M. and Burles, Scott},
   year={2000},
   month=May }

@article{Serpico_2004,
doi = {10.1088/1475-7516/2004/12/010},
url = {https://doi.org/10.1088/1475-7516/2004/12/010},
year = {2004},
month = {dec},
publisher = {},
volume = {2004},
number = {12},
pages = {010},
author = {Serpico, P D and Esposito, S and Iocco, F and Mangano, G and Miele, G and Pisanti, O},
title = {Nuclear reaction network for primordial nucleosynthesis: a detailed analysis of rates,
uncertainties and light nuclei yields},
journal = {Journal of Cosmology and Astroparticle Physics},
}

@INPROCEEDINGS{coc_2010,
       author = {{Coc}, Alain and {Vangioni}, Elisabeth},
        title = "{Big-Bang nucleosynthesis with updated nuclear data}",
    booktitle = {Journal of Physics Conference Series},
         year = 2010,
       series = {Journal of Physics Conference Series},
       volume = {202},
        month = jan,
    publisher = {IOP},
          eid = {012001},
        pages = {012001},
          doi = {10.1088/1742-6596/202/1/012001},
       adsurl = {https://ui.adsabs.harvard.edu/abs/2010JPhCS.202a2001C}
}

@article{Cyburt_2016,
  title = {Big bang nucleosynthesis: Present status},
  author = {Cyburt, Richard H. and Fields, Brian D. and Olive, Keith A. and Yeh, Tsung-Han},
  journal = {Rev. Mod. Phys.},
  volume = {88},
  issue = {1},
  pages = {015004},
  numpages = {22},
  year = {2016},
  month = {Feb},
  publisher = {American Physical Society},
  doi = {10.1103/RevModPhys.88.015004},
  url = {https://link.aps.org/doi/10.1103/RevModPhys.88.015004}
}

@article{Fields_2020,
   title={Big-Bang Nucleosynthesis after Planck},
   volume={2020},
   ISSN={1475-7516},
   url={http://dx.doi.org/10.1088/1475-7516/2020/03/010},
   DOI={10.1088/1475-7516/2020/03/010},
   number={03},
   journal={Journal of Cosmology and Astroparticle Physics},
   publisher={IOP Publishing},
   author={Fields, Brian D. and Olive, Keith A. and Yeh, Tsung-Han and Young, Charles},
   year={2020},
   month=Mar, pages={010–010} }

@article{Nollett_2014,
   title={{BBN and the CMB constrain light, electromagnetically coupled WIMPs}},
   volume={89},
   ISSN={1550-2368},
   url={http://dx.doi.org/10.1103/PhysRevD.89.083508},
   DOI={10.1103/physrevd.89.083508},
   number={8},
   journal={Physical Review D},
   publisher={American Physical Society (APS)},
   author={Nollett, Kenneth M. and Steigman, Gary},
   year={2014},
   month=Apr }

@article{DESI:2025zgx,
    author = "Abdul Karim, M. and others",
    collaboration = "DESI",
    title = "{DESI DR2 results. II. Measurements of baryon acoustic oscillations and cosmological constraints}",
    eprint = "2503.14738",
    archivePrefix = "arXiv",
    primaryClass = "astro-ph.CO",
    reportNumber = "FERMILAB-PUB-25-0169-PPD",
    doi = "10.1103/tr6y-kpc6",
    journal = "Phys. Rev. D",
    volume = "112",
    number = "8",
    pages = "083515",
    year = "2025"
}

@article{DESI:2025qqy,
    author = "Andrade, U. and others",
    collaboration = "DESI",
    title = "{Validation of the DESI DR2 measurements of baryon acoustic oscillations from galaxies and quasars}",
    eprint = "2503.14742",
    archivePrefix = "arXiv",
    primaryClass = "astro-ph.CO",
    reportNumber = "FERMILAB-PUB-25-0162-PPD",
    doi = "10.1103/kdys-w8vl",
    journal = "Phys. Rev. D",
    volume = "112",
    number = "8",
    pages = "083512",
    year = "2025"
}

@misc{ganguly2026consistentnrmefffitting,
      title={Consistent $N_{\rm eff}$ fitting in big bang nucleosynthesis analysis}, 
      author={Sougata Ganguly and Tae Hyun Jung and Seokhoon Yun},
      year={2026},
      eprint={2507.23354},
      archivePrefix={arXiv},
      primaryClass={hep-ph},
      url={https://arxiv.org/abs/2507.23354}, 
}

@article{Giovanetti_2022,
   title={Joint Cosmic Microwave Background and Big Bang Nucleosynthesis Constraints on Light Dark Sectors with Dark Radiation},
   volume={129},
   ISSN={1079-7114},
   url={http://dx.doi.org/10.1103/PhysRevLett.129.021302},
   DOI={10.1103/physrevlett.129.021302},
   number={2},
   journal={Physical Review Letters},
   publisher={American Physical Society (APS)},
   author={Giovanetti, Cara and Lisanti, Mariangela and Liu, Hongwan and Ruderman, Joshua T.},
   year={2022},
   month=July }

@article{Giovanetti_2025,
   title={Neutrino-dark sector equilibration and primordial element abundances},
   volume={111},
   ISSN={2470-0029},
   url={http://dx.doi.org/10.1103/PhysRevD.111.043526},
   DOI={10.1103/physrevd.111.043526},
   number={4},
   journal={Physical Review D},
   publisher={American Physical Society (APS)},
   author={Giovanetti, Cara and Schmaltz, Martin and Weiner, Neal},
   year={2025},
   month=Feb }

@article{Escudero_Abenza_2026,
   title={{Nucleosynthesis and CMB bounds on photophilic ALPs: a fresh look}},
   volume={86},
   ISSN={1434-6052},
   url={http://dx.doi.org/10.1140/epjc/s10052-026-15544-z},
   DOI={10.1140/epjc/s10052-026-15544-z},
   number={5},
   journal={The European Physical Journal C},
   publisher={Springer Science and Business Media LLC},
   author={Escudero Abenza, Miguel and Garcia-Perez, Clara and Ovchynnikov, Maksym},
   year={2026},
   month=May }

@article{Sabti_2020,
   title={{Refined bounds on MeV-scale thermal dark sectors from BBN and the CMB}},
   volume={2020},
   ISSN={1475-7516},
   url={http://dx.doi.org/10.1088/1475-7516/2020/01/004},
   DOI={10.1088/1475-7516/2020/01/004},
   number={01},
   journal={Journal of Cosmology and Astroparticle Physics},
   publisher={IOP Publishing},
   author={Sabti, Nashwan and Alvey, James and Escudero, Miguel and Fairbairn, Malcolm and Blas, Diego},
   year={2020},
   month=Jan, pages={004–004} }

@article{Li_2020,
   title={Probing dark photons in the early universe with big bang nucleosynthesis},
   volume={2020},
   ISSN={1475-7516},
   url={http://dx.doi.org/10.1088/1475-7516/2020/12/049},
   DOI={10.1088/1475-7516/2020/12/049},
   number={12},
   journal={Journal of Cosmology and Astroparticle Physics},
   publisher={IOP Publishing},
   author={Li, Jung-Tsung and Fuller, George M. and Grohs, Evan},
   year={2020},
   month=Dec, pages={049–049} }

@article{Barbieri_2025,
   title={Current Constraints on Cosmological Scenarios with Very Low Reheating Temperatures},
   volume={135},
   ISSN={1079-7114},
   url={http://dx.doi.org/10.1103/j5rj-dz1k},
   DOI={10.1103/j5rj-dz1k},
   number={18},
   journal={Physical Review Letters},
   publisher={American Physical Society (APS)},
   author={Barbieri, Nicola and Brinckmann, Thejs and Gariazzo, Stefano and Lattanzi, Massimiliano and Pastor, Sergio and Pisanti, Ofelia},
   year={2025},
   month=Oct }

@article{de_Salas_2015,
   title={Bounds on very low reheating scenarios after {P}lanck},
   volume={92},
   ISSN={1550-2368},
   url={http://dx.doi.org/10.1103/PhysRevD.92.123534},
   DOI={10.1103/physrevd.92.123534},
   number={12},
   journal={Physical Review D},
   publisher={American Physical Society (APS)},
   author={de Salas, P. F. and Lattanzi, M. and Mangano, G. and Miele, G. and Pastor, S. and Pisanti, O.},
   year={2015},
   month=Dec }

@article{Hasegawa_2019,
   title={{MeV}-scale reheating temperature and thermalization of oscillating neutrinos by radiative and hadronic decays of massive particles},
   volume={2019},
   ISSN={1475-7516},
   url={http://dx.doi.org/10.1088/1475-7516/2019/12/012},
   DOI={10.1088/1475-7516/2019/12/012},
   number={12},
   journal={Journal of Cosmology and Astroparticle Physics},
   publisher={IOP Publishing},
   author={Hasegawa, Takuya and Hiroshima, Nagisa and Kohri, Kazunori and Hansen, Rasmus S.L. and Tram, Thomas and Hannestad, Steen},
   year={2019},
   month=Dec, pages={012–012} }

@article{Millea_2015,
   title={New bounds for axions and axion-like particles with {keV-GeV} masses},
   volume={92},
   ISSN={1550-2368},
   url={http://dx.doi.org/10.1103/PhysRevD.92.023010},
   DOI={10.1103/physrevd.92.023010},
   number={2},
   journal={Physical Review D},
   publisher={American Physical Society (APS)},
   author={Millea, Marius and Knox, Lloyd and Fields, Brian D.},
   year={2015},
   month=July }

@article{Depta_2020,
   title={Robust cosmological constraints on axion-like particles},
   volume={2020},
   ISSN={1475-7516},
   url={http://dx.doi.org/10.1088/1475-7516/2020/05/009},
   DOI={10.1088/1475-7516/2020/05/009},
   number={05},
   journal={Journal of Cosmology and Astroparticle Physics},
   publisher={IOP Publishing},
   author={Depta, Paul Frederik and Hufnagel, Marco and Schmidt-Hoberg, Kai},
   year={2020},
   month=May, pages={009–009} }

@article{Berlin_2019,
   title={Dark sector equilibration during nucleosynthesis},
   volume={100},
   ISSN={2470-0029},
   url={http://dx.doi.org/10.1103/PhysRevD.100.015038},
   DOI={10.1103/physrevd.100.015038},
   number={1},
   journal={Physical Review D},
   publisher={American Physical Society (APS)},
   author={Berlin, Asher and Blinov, Nikita and Li, Shirley Weishi},
   year={2019},
   month=July }

@article{Jung_2026,
   title={New bounds on heavy {QCD} axions from big bang nucleosynthesis},
   volume={113},
   ISSN={2470-0029},
   url={http://dx.doi.org/10.1103/l2m1-h1cp},
   DOI={10.1103/l2m1-h1cp},
   number={5},
   journal={Physical Review D},
   publisher={American Physical Society (APS)},
   author={Jung, Tae Hyun and Okui, Takemichi and Tobioka, Kohsaku and Wang, Jiabao},
   year={2026},
   month=Mar }

@article{Serpico_2004b,
   title={{MeV}-mass dark matter and primordial nucleosynthesis},
   volume={70},
   ISSN={1550-2368},
   url={http://dx.doi.org/10.1103/PhysRevD.70.043526},
   DOI={10.1103/physrevd.70.043526},
   number={4},
   journal={Physical Review D},
   publisher={American Physical Society (APS)},
   author={Serpico, Pasquale D. and Raffelt, Georg G.},
   year={2004},
   month=Aug }

@article{Ho_2013,
   title={Sterile neutrinos and light dark matter save each other},
   volume={87},
   ISSN={1550-2368},
   url={http://dx.doi.org/10.1103/PhysRevD.87.065016},
   DOI={10.1103/physrevd.87.065016},
   number={6},
   journal={Physical Review D},
   publisher={American Physical Society (APS)},
   author={Ho, Chiu Man and Scherrer, Robert J.},
   year={2013},
   month=Mar }

@article{Berezhiani_2013,
   title={{BBN} with light dark matter},
   volume={2013},
   ISSN={1475-7516},
   url={http://dx.doi.org/10.1088/1475-7516/2013/02/010},
   DOI={10.1088/1475-7516/2013/02/010},
   number={02},
   journal={Journal of Cosmology and Astroparticle Physics},
   publisher={IOP Publishing},
   author={Berezhiani, Zurab and Dolgov, Aleksander and Tkachev, Igor},
   year={2013},
   month=Feb, pages={010–010} }

@article{Cadamuro_2012,
   title={Cosmological bounds on pseudo Nambu-Goldstone bosons},
   volume={2012},
   ISSN={1475-7516},
   url={http://dx.doi.org/10.1088/1475-7516/2012/02/032},
   DOI={10.1088/1475-7516/2012/02/032},
   number={02},
   journal={Journal of Cosmology and Astroparticle Physics},
   publisher={IOP Publishing},
   author={Cadamuro, Davide and Redondo, Javier},
   year={2012},
   month=Feb, pages={032–032} }

@article{Blas:2011rf,
    author = "Blas, Diego and Lesgourgues, Julien and Tram, Thomas",
    title = "{The Cosmic Linear Anisotropy Solving System (CLASS) II: Approximation schemes}",
    eprint = "1104.2933",
    archivePrefix = "arXiv",
    primaryClass = "astro-ph.CO",
    reportNumber = "CERN-PH-TH-2011-082, LAPTH-010-11",
    doi = "10.1088/1475-7516/2011/07/034",
    journal = "JCAP",
    volume = "07",
    pages = "034",
    year = "2011"
}

@misc{sharma2026recoupleddarkradiationreconciling,
      title={{Recoupled Dark Radiation reconciling CMB and DESI BAO measurements}}, 
      author={Ravi Kumar Sharma and Maria Archidiacono and Julien Lesgourgues},
      year={2026},
      eprint={2605.18716},
      archivePrefix={arXiv},
      primaryClass={astro-ph.CO},
      url={https://arxiv.org/abs/2605.18716}, 
}

@misc{garciaquintero2025cosmologicalimplicationsdesidr2,
      title={{Cosmological implications of DESI DR2 BAO measurements in light of the latest ACT DR6 CMB data}}, 
      author={C. Garcia-Quintero and others},
      year={2025},
      eprint={2504.18464},
      archivePrefix={arXiv},
      primaryClass={astro-ph.CO},
      url={https://arxiv.org/abs/2504.18464}, 
}

@article{Torrado:2020dgo,
    author = "Torrado, Jesus and Lewis, Antony",
    title = "{Cobaya: Code for Bayesian Analysis of hierarchical physical models}",
    eprint = "2005.05290",
    archivePrefix = "arXiv",
    primaryClass = "astro-ph.IM",
    reportNumber = "TTK-20-15",
    doi = "10.1088/1475-7516/2021/05/057",
    journal = "JCAP",
    volume = "05",
    pages = "057",
    year = "2021"
}

@software{2019ascl.soft10019T,
       author = {{Torrado}, Jes{\'u}s and {Lewis}, Antony},
        title = "{Cobaya: Bayesian analysis in cosmology}",
 howpublished = {Astrophysics Source Code Library, record ascl:1910.019},
         year = 2019,
        month = oct,
          eid = {ascl:1910.019},
archivePrefix = {ascl},
       eprint = {1910.019},
       adsurl = {https://ui.adsabs.harvard.edu/abs/2019ascl.soft10019T}
}

@article{Czarnecki:2018okw,
    author = "Czarnecki, Andrzej and Marciano, William J. and Sirlin, Alberto",
    title = "{Neutron Lifetime and Axial Coupling Connection}",
    eprint = "1802.01804",
    archivePrefix = "arXiv",
    primaryClass = "hep-ph",
    reportNumber = "ALBERTA-THY-1-18",
    doi = "10.1103/PhysRevLett.120.202002",
    journal = "Phys. Rev. Lett.",
    volume = "120",
    number = "20",
    pages = "202002",
    year = "2018"
}

@article{abcmb,
doi = {10.1088/1475-7516/2026/08/078},
url = {https://doi.org/10.1088/1475-7516/2026/08/078},
year = {2026},
month = {aug},
publisher = {IOP Publishing},
volume = {2026},
number = {08},
pages = {078},
author = {Zhou, Zilu and Giovanetti, Cara and Liu, Hongwan},
title = {{ABCMB: A Python+JAX Package for the Cosmic Microwave Background Power Spectrum}},
journal = {Journal of Cosmology and Astroparticle Physics}
}

@article{Schulte1972,
	author = "R. L. Schulte and M. Cosack and A. W. Obst and J. L. Weil",
	title = "{``$^2$H+ reactions from 1.96 to 6.20 MeV''}",
	journal = "Nucl. Phys. A",
	volume = "192",
	pages = "609-624",
	year = "1972",
	doi = "10.1016/0375-9474(72)90093-0"
}

@article{Tumino2014,
	author = "Aurora {Tumino \textit{et al.}}",
	title = "{``New determination of the $^2$H($d$,$p$)$^3$H and $^2$H($d$,$n$)$^3$He reaction rates at astrophysical energies''}",
	journal = "Astrophys. J.",
	volume = "785",
	pages = "96",
	year = "2014",
	doi = "10.1088/0004-637X/785/2/96"
}

@article{Mossa_2020,
	Author = {Mossa, V. and Sothers},
	Journal = {Nature},
	Number = {7833},
	Pages = {210--213},
	Title = {The baryon density of the Universe from an improved rate of deuterium burning},
	Volume = {587},
	Year = {2020}}

@article{Yeh_2022,
   title={{Probing physics beyond the standard model:  limits from BBN and the CMB independently and combined},
   volume={2022}},
   ISSN={1475-7516},
   url={http://dx.doi.org/10.1088/1475-7516/2022/10/046},
   DOI={10.1088/1475-7516/2022/10/046},
   number={10},
   journal={Journal of Cosmology and Astroparticle Physics},
   publisher={IOP Publishing},
   author={Yeh, Tsung-Han and Shelton, Jessie and Olive, Keith A. and Fields, Brian D.},
   year={2022},
   month=Oct, pages={046} }

@article{Arai2011,
  title = {{Tensor Force Manifestations in Ab Initio Study of the $^{2}\mathrm{H}(d,\ensuremath{\gamma})^{4}\mathrm{He}$, $^{2}\mathrm{H}(d,p)^{3}\mathrm{H}$, and $^{2}\mathrm{H}(d,n)^{3}\mathrm{He}$ Reactions}},
  author = {Arai, K. and Aoyama, S. and Suzuki, Y. and Descouvemont, P. and Baye, D.},
  journal = {Phys. Rev. Lett.},
  volume = {107},
  issue = {13},
  pages = {132502},
  numpages = {5},
  year = {2011},
  month = {Sep},
  publisher = {American Physical Society},
  doi = {10.1103/PhysRevLett.107.132502},
  url = {https://link.aps.org/doi/10.1103/PhysRevLett.107.132502}
}

@article{Gomez_Inesta_2017,
   title={Bayesian Estimation of Thermonuclear Reaction Rates for Deuterium+Deuterium Reactions},
   volume={849},
   ISSN={1538-4357},
   url={http://dx.doi.org/10.3847/1538-4357/aa9025},
   DOI={10.3847/1538-4357/aa9025},
   number={2},
   journal={The Astrophysical Journal},
   publisher={American Astronomical Society},
   author={G\'omez I\~nesta, \'A. and Iliadis, C. and Coc, A.},
   year={2017},
   month=Nov, pages={134} }

@article{Moscoso_2021,
   title={{Bayesian Estimation of the D(p,$\gamma$)$^3$He Thermonuclear Reaction Rate}},
   volume={923},
   ISSN={1538-4357},
   url={http://dx.doi.org/10.3847/1538-4357/ac1db0},
   DOI={10.3847/1538-4357/ac1db0},
   number={1},
   journal={The Astrophysical Journal},
   publisher={American Astronomical Society},
   author={Moscoso, Joseph and de Souza, Rafael S. and Coc, Alain and Iliadis, Christian},
   year={2021},
   month=Dec, pages={49} }

@article{Coc_2015,
   title={{New reaction rates for improved primordial D/H calculation and the cosmic evolution of deuterium}},
   volume={92},
   ISSN={1550-2368},
   url={http://dx.doi.org/10.1103/PhysRevD.92.123526},
   DOI={10.1103/physrevd.92.123526},
   number={12},
   journal={Physical Review D},
   publisher={American Physical Society (APS)},
   author={Coc, Alain and Petitjean, Patrick and Uzan, Jean-Philippe and Vangioni, Elisabeth and Descouvemont, Pierre and Iliadis, Christian and Longland, Richard},
   year={2015},
   month=Dec }

@misc{schoneberg_2024,
      title={{The 2024 BBN baryon abundance update}}, 
      author={Nils Sch\"oneberg},
      year={2026},
      eprint={2401.15054},
      archivePrefix={arXiv},
      primaryClass={astro-ph.CO},
      url={https://arxiv.org/abs/2401.15054}, 
}

@article{CosmoVerseNetwork:2025alb,
    author = "Di Valentino, Eleonora and others",
    collaboration = "CosmoVerse Network",
    title = "{The CosmoVerse White Paper: Addressing observational tensions in cosmology with systematics and fundamental physics}",
    eprint = "2504.01669",
    archivePrefix = "arXiv",
    primaryClass = "astro-ph.CO",
    doi = "10.1016/j.dark.2025.101965",
    journal = "Phys. Dark Univ.",
    volume = "49",
    pages = "101965",
    year = "2025"
}

@article{Shah:2021onj,
    author = "Shah, Paul and Lemos, Pablo and Lahav, Ofer",
    title = "{A buyer{\textquoteright}s guide to the Hubble constant}",
    eprint = "2109.01161",
    archivePrefix = "arXiv",
    primaryClass = "astro-ph.CO",
    doi = "10.1007/s00159-021-00137-4",
    journal = "Astron. Astrophys. Rev.",
    volume = "29",
    number = "1",
    pages = "9",
    year = "2021"
}

@article{Burns_2023,
   title={Indications for a Nonzero Lepton Asymmetry from Extremely Metal-Poor Galaxies},
   volume={130},
   ISSN={1079-7114},
   url={http://dx.doi.org/10.1103/PhysRevLett.130.131001},
   DOI={10.1103/physrevlett.130.131001},
   number={13},
   journal={Physical Review Letters},
   publisher={American Physical Society (APS)},
   author={Burns, Anne-Katherine and Tait, Tim M. P. and Valli, Mauro},
   year={2023},
   month=Mar }

@article{DESI:2025zpo,
    author = "Abdul Karim, M. and others",
    collaboration = "DESI",
    title = "{DESI DR2 results. I. Baryon acoustic oscillations from the Lyman alpha forest}",
    eprint = "2503.14739",
    archivePrefix = "arXiv",
    primaryClass = "astro-ph.CO",
    reportNumber = "FERMILAB-PUB-25-0167-PPD",
    doi = "10.1103/2wwn-xjm5",
    journal = "Phys. Rev. D",
    volume = "112",
    number = "8",
    pages = "083514",
    year = "2025"
}

@article{Akita_2025a,
   title={New Physics Decaying into Metastable Particles: Impact on Cosmic Neutrinos},
   volume={134},
   ISSN={1079-7114},
   url={http://dx.doi.org/10.1103/PhysRevLett.134.121001},
   DOI={10.1103/physrevlett.134.121001},
   number={12},
   journal={Physical Review Letters},
   publisher={American Physical Society (APS)},
   author={Akita, Kensuke and Baur, Gideon and Ovchynnikov, Maksym and Schwetz, Thomas and Syvolap, Vsevolod},
   year={2025},
   month=Mar }

@article{Akita_2025b,
   title={{Dynamics of metastable standard model particles from long-lived particle decays in the MeV primordial plasma}},
   volume={111},
   ISSN={2470-0029},
   url={http://dx.doi.org/10.1103/PhysRevD.111.063542},
   DOI={10.1103/physrevd.111.063542},
   number={6},
   journal={Physical Review D},
   publisher={American Physical Society (APS)},
   author={Akita, Kensuke and Baur, Gideon and Ovchynnikov, Maksym and Schwetz, Thomas and Syvolap, Vsevolod},
   year={2025},
   month=Mar }

\end{document}